\documentclass[lettersize,journal]{IEEEtran}
\usepackage{amsmath,amsfonts}
\usepackage{algorithmic}
\usepackage{algorithm}
\usepackage{array}
\usepackage{textcomp}
\usepackage{stfloats}
\usepackage{url}

\usepackage{verbatim}
\usepackage{graphicx}

\usepackage{threeparttable}
\usepackage{multirow}
\usepackage{soul}
\usepackage{color, xcolor}

\usepackage{hyperref}
\hypersetup{
hidelinks
}
\usepackage{amsmath,epsfig,amssymb,subfigure,bm,dsfont}
\usepackage[square, comma, sort&compress, numbers]{natbib} 
\usepackage{bookmark}

\begin{document}
	
\sethlcolor{yellow}


\title{Wafer-scale Computing: Advancements, Challenges, and Future Perspectives}


\author{Yang Hu, \IEEEmembership{Member,~IEEE}, Xinhan Lin, 
    Huizheng Wang, Zhen He, \\
    Xingmao Yu, Jiahao Zhang, Qize Yang, Zheng Xu, \\
    Sihan Guan, Jiahao Fang, Haoran Shang, Xinru Tang, Xu Dai, \\
    Shaojun Wei, \IEEEmembership{Fellow,~IEEE},
	and Shouyi Yin, \IEEEmembership{Senior~Member,~IEEE}

\thanks{
Yang Hu, Huizheng Wang, Zhen He, Xingmao Yu, Jiahao Zhang, Qize Yang, Zheng Xu, Sihan Guan, Jiahao Fang, Haoran Shang, Xinru Tang, Shaojun Wei, and Shouyi Yin 
are with the School of Integrated Circuits, Tsinghua University, Beijing 100084, China.

Xinhan Lin and Xu Dai are with the Shanghai Artificial Intelligence Laboratory, Shanghai 200232, China.

Corresponding Author: Shouyi Yin (yinsy@tsinghua.edu.cn)}
}

\maketitle

\begin{abstract}

Nowadays, artificial intelligence (AI) technology with large models plays an increasingly important role in both academia and industry.
It also brings a rapidly increasing demand for the computing power of the hardware. As the computing demand for AI continues to grow, the growth of hardware computing power has failed to keep up. This has become a significant factor restricting the development of AI. 
The augmentation of hardware computing power is mainly propelled by the escalation of transistor density and chip area. However, the former is impeded by the termination of the Moore's Law and Dennard scaling, and the latter is significantly restricted by the challenge of disrupting the legacy fabrication equipment and process. 

In recent years, advanced packaging technologies that have gradually matured are increasingly used to implement bigger chips that integrate multiple chiplets, while still providing interconnections with chip-level density and bandwidth. This technique points out a new path of continuing the increase of computing power while leveraging the current fabrication process without significant disruption. Enabled by this technique, a chip can extend to a size of wafer-scale (over 10,000 mm$^2$), provisioning orders of magnitude more computing capabilities (several POPS within just one monolithic chip) and die-to-die bandwidth density (over 15 GB/s/mm) than regular chips, and emerges a new Wafer-scale Computing paradigm.
Compared to conventional high-performance computing paradigms such as multi-accelerator and datacenter-scale computing, Wafer-scale Computing shows remarkable advantages in communication bandwidth, integration density, and programmability potential.
Not surprisingly, disruptive Wafer-scale Computing also brings unprecedented design challenges for hardware architecture, design-\textbackslash system- technology co-optimization, power and cooling systems, and compiler tool chain.
At present, there are no comprehensive surveys summarizing the current state and design insights of Wafer-scale Computing. This paper aims to take the first step to help academia and industry review existing wafer-scale chips and essential technologies in a one-stop manner. So that people can conveniently grasp the basic knowledge and key points, understand the achievements and shortcomings of existing research, and contribute to this promising research direction. 

\end{abstract}



\section{Introduction\label{secInt}}

In today's world, artificial intelligence (AI) technology with large models plays an increasingly important role in promoting scientific progress and is becoming a basic tool for human beings to explore the world.
For example, 
AlphaFold2 can predict 98.5\% of human protein sequences with high confidence while the traditional method can only cover 17\% \cite{Alphafold},
AlphaTensor discovers fast matrix multiplication algorithms and surpasses the classic algorithm discovered 50 years ago for the first time \cite{AlphaTensor},
DGMR beats competitive methods in 89\% of the cases of two-hour-ahead high-resolution weather forecasting \cite{DGMR}, and
DM21 performs better than traditional functionals for describing matter at the quantum level \cite{DM21}. As an AI language model developed by OpenAI, ChatGPT \cite{chatgpt} has gained significant popularity and recognition in the field of natural language processing (NLP) since its release.

\begin{figure}
	\centering
	\includegraphics[width=.95\linewidth]{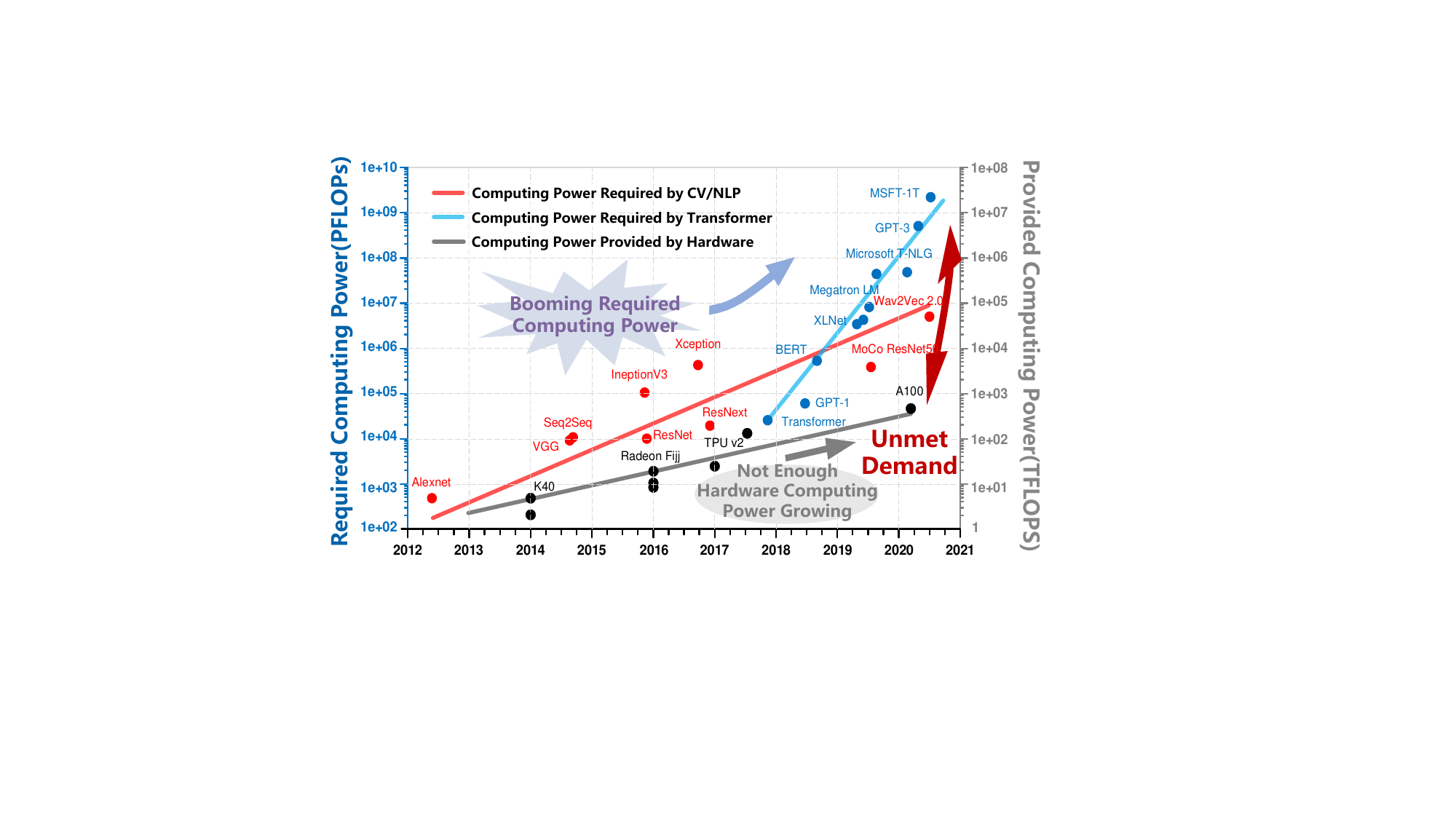}
	\caption{Scissors difference between the required/provided computing powers.}
	\label{figIntroSciDiff}
\end{figure}

Although large AI models offer numerous benefits to research, production, and daily life, they also impose an immense demand on hardware computing power.
With the emergence of the transformer model\cite{NIPS17Transformer} in recent years, the demand for computing power required by large models has experienced explosive growth, increasing by a factor of 1,000 within a span of two years, as shown in Figure \ref{figIntroSciDiff}.
By contrast, we observe that the growth rate of hardware computing power utilized to train large models only doubles over a period of two years. As a result, a significant gap exists between the computing power demanded by large models and that which is currently available from chips, which serves as a key limiting factor for the advancement of AI.

\begin{figure*}
	\centering
	\includegraphics[width=.95\linewidth]{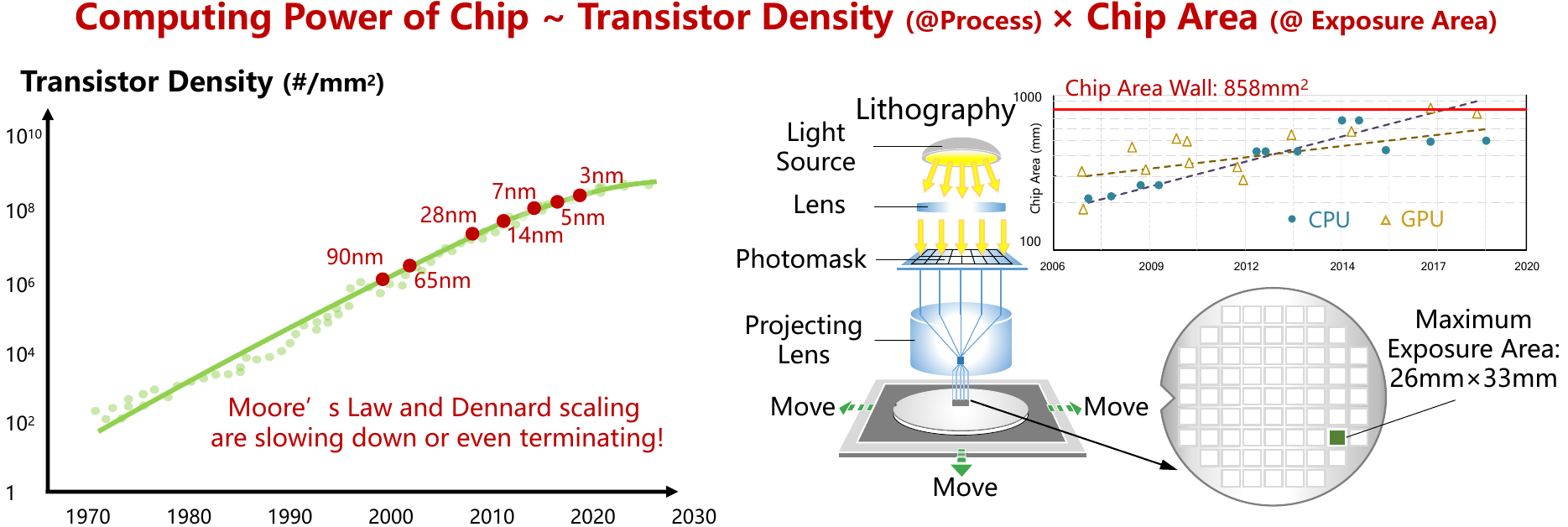}
	\caption{Limiting factors of computing power.}
	\label{figIntroChipCompPow}
\end{figure*}

The computing power of a chip can be represented as the total number of transistors integrated within it, which is the product of transistor density (i.e., the number of transistors per unit area) and area. Transistor density is mainly determined by the advancement of the chip process, whereas chip area is predominantly determined by the caliber of the chip photolithography process, and is limited by a maximum reticle size. Unfortunately, both paths are presently encountering significant obstacles that impede their continuation.

Firstly, the increasing difficulty of improving transistor density due to the slowing or termination of Moore's Law and Dennard scaling has become a significant issue after 3nm \cite{theis2017end}. Secondly, the chip area is generally restricted by reticle size (i.e., the largest area of the wafer that can be patterned using a lithography stepper system), which is challenging to enlarge while maintaining the most up-to-date chip process and preserving yield \cite{MICRO21CBStory}.

In recent years, advanced packaging technologies that have gradually matured are increasingly used to implement bigger chips that integrate multiple chiplets/dielets, while still providing interconnections with chip-level density and bandwidth \cite{HCS19Cerebras,rocki2020fast,Online21TeslaDojo}. This points out a new path of continuing the increase of computing power while leveraging the current fabrication process without significant disruption,
and emerges a new and exciting \textbf{\textit{Wafer-scale Computing}} paradigm.
In this paper, Wafer-scale Computing is defined as the computing paradigm which extends the chip area to a size of wafer-scale (over 10,000 mm$^2$) through leveraging the advanced packaging \cite{Book21AdvPack} or field stitching \cite{ICWSI92Lithographic} techniques to tightly integrate multiple chiplets/dielets, provisioning orders of magnitude more computing capabilities (several POPS within just one monolithic chip) and die-to-die bandwidth density (over 15 GB/s/mm) than regular chips, while leveraging the current fabrication process without significant disruption.


One might question why not construct an accelerator cluster (such as NVIDIA's GPU cluster \cite{Online22NVIDIAH100}) using multiple chips to achieve computing power that surpasses that of a single conventional chip. Alternatively, what benefits do wafer-scale chips have over accelerator clusters? 
The answers mainly include the following aspects:
\begin{itemize}
	\item One of the primary advantages of wafer-level integration is the significant enhancement of die-to-die bandwidth. This benefit is straightforward and evident. For instance, NVIDIA's interconnect provides a bandwidth of 900GB/s for H100 GPUs \cite{Online22NVIDIAH100}, whereas every D1 die edge in Tesla Dojo delivers a bandwidth of 2TB/s \cite{HCS22Dojo}. This high bandwidth breaks the previous memory limitations and opens up new design spaces for exploring more aggressive application solutions that can fully utilize the computing power available.
	\item Moreover, wafer-scale chips offer better integration density, which implies higher size and form factor efficiencies. For instance, a Tesla Dojo training tile can tightly integrate 25 regular-sized dies, whereas 25 NVIDIA H100s require a complete package for each GPU and take up more than ten times the total area. This size advantage may be preferred in supercomputing applications with limited space, such as aerospace and military applications.
	\item Lastly, wafer-scale chips offer significant potential for programmability. Compared to GPU clusters, wafer-scale chips have considerably less overhead when it comes to inter-die and intra-die data communication, and current wafer-scale chip designs \cite{2048chiplet,HCS22Dojo,HCS22Cerebras} strive to minimize this gap even further. This means that programmers do not have to be overly concerned about data access across multiple dies, and compilers can generate more flexible and fine-grained hardware resource partitioning and task mapping to more effectively utilize the computing power available.
\end{itemize}

Not surprisingly, this disruptive wafer-scale
computing paradigm also brings unprecedented challenges in nearly every design aspect.
\begin{itemize}
	\item To start, the hardware footprints for Wafer-scale Computing system is orders of magnitudes larger than traditional chips, bringing significantly larger design spaces, which nullifies conventional architectural design methodologies.
	\item From an architectural perspective, the current hardware execution model exhibits inadequate scalability. To enable the efficient operation of Wafer-scale Computing systems, it is imperative to develop new computational and execution models. 
 	\item From an integration/packaging perspective, the current advanced packaging technologies are being implemented in Wafer-scale Computing only as experimental runs and in an ad hoc manner. There is still a need for systematic design principles to determine the appropriate wafer-scale substrate techniques and layout, yield issues, and to co-optimize the packaging and system design.
	\item From a systems perspective, there is currently no cross-stack system design methodology for Wafer-scale Computing systems. Such systems operate comprehensively, with tightly coupled computing dies, wafer-scale substrates, power and cooling systems, and mechanical parts. The performance and availability of any single component can significantly affect the other parts.
 	\item From a software perspective, the current software stack has not encountered computing resources on the scale of Wafer-scale Computing and lacks efficient mechanisms to fully leverage the power of such systems. There is a high demand for compiling and execution mechanisms to map big AI models to Wafer-scale Computing resources and run them efficiently. 
\end{itemize}

At present, there are no comprehensive surveys summarizing the current states and design insights of Wafer-scale Computing. 
The objective of this paper is to provide a comprehensive overview of existing wafer-scale chips and their essential technologies in a single source. This will enable individuals in academia and industry to easily comprehend the fundamental concepts and key aspects, assess the accomplishments and limitations of existing research, and make contributions to this promising research field.

The rest of this paper is organized as follows: 
Section \ref{secBkg} introduces the background technologies.
Section \ref{secArch} discusses the key architecture design points of wafer-scale chips based on existing works.
Section \ref{secInterconnect} reviews common-used interconnect interfaces and protocols which may be used in wafer-scale chips.
Section \ref{secTool} 
reviews typical compiler tools for driving large-scale acceleration platforms (including traditional ones and Wafer-scale computing ones), and discusses what make the compilation of Wafer-scale Computing different from traditional one.
Section \ref{secInteg} reviews how existing works integrate wafer-scale chips.
Section \ref{secSystem} discusses how to deliver power and clock to the whole wafer-scale chip, and how to solve the cooling problem under such high integration density.
Section \ref{secFault} provides a summary of fault tolerance designs.
Section \ref{secApp} presents a brief introduction to the important scientific computing applications of Wafer-scale Computing outside of AI computing domain.
Section \ref{secCon} concludes the paper.


\section{Background\label{secBkg}}

In this section, we will introduce the efforts for improving the computing power of AI acceleration.

\subsection{Conventional Accelerators for AI Tasks\label{secBkgAcc}}

As Moore’s Law and Dennard scaling are slowing down or even terminating, various accelerators different from classic Von Neumann architectures have been proposed to meet the ever-increasing user demand of AI applications.
These accelerators mainly include general-purpose graphics processing units (GPGPUs) \cite{GPGPU}, field-programmable gate arrays (FPGAs) \cite{FPGA}, application-specific integrated circuits (ASICs)\cite{ASIC} and coarse-grained reconfigurable architectures (CGRAs) \cite{CGRA}.
GPGPUs, such as NVIDIA's A100 \cite{Online20NVIDIAA100} and H100 \cite{Online22NVIDIAH100}, are designed for scientific computation, encryption/decryption, and, of course, AI computation.
GPGPUs have much more computing units than Von Neumann CPUs, and adopt the supporting execution mechanism (e.g., single instruction multiple thread, SIMT \cite{Book12SIMT}),
so they can surpass CPUs in the performance of high parallel computing.
Since GPGPUs usually have large volume and power consumption, they are mainly used in data centers.
With the popularization of AI and increasingly diversified user needs, researchers also develop more domain-specific accelerators on the platforms other than GPGPUs, to meet the demands such as energy efficiency and portability.
These accelerators, including FPGAs, ASICs and CGRAs, simplify the computing units, data-paths and memories to exactly match the application scenarios, so they usually have the advantages of energy efficiency and area efficiency over GPGPUs.
FPGAs offer fine-grained programmable logic, computing and storage units, for allowing users to totally customize the computing path structure according to algorithm requirements.
CGRAs offer coarse-grained reconfigurability to provide flexibility limited but sufficient for the target application scenarios.
ASICs cannot change its functions in the original sense, however, for AI accelerators the boundary of ASIC and CGRA is blurred, many so-called ASIC-based AI accelerators (e.g., \cite{kwon2018maeri,kim20192,chen2019eyeriss} ) also have coarse-grained reconfigurability. 

Those mentioned above accelerators have in common that they adopt 1) large numbers of parallel computing units to provide high computing power, 2) hierarchical memory systems to reuse the data and reduce the data movement overhead, 3) convenient networks on chip (NoCs) to connect the distributed computing units and memories and 4) elaborate dataflow to utilize as more computing power as possible.
Designers make decisions on these issues, to strive for the best performance, efficiency or flexibility.

\subsection{Accelerator Clusters \label{secBkgMultiChip}}

Caused by the slowdown of CMOS scaling and the limit of chip area, the number of available transistors on a monolithic chip becomes stagnant nowadays, which will seriously restrict the processor design optimization space.
The gains from specialized architecture design will gradually diminish and ultimately hit an upper-bound, which is called accelerator wall \cite{AccWall}.
To address the gap between the diminishing gains of accelerator performance and the growing demand of computing power for AI tasks, a straightforward way is to group multiple individual accelerators into a cluster, so that much more computing power can be exploited.

\begin{figure}
	\centering
	\includegraphics[width=.95\linewidth]{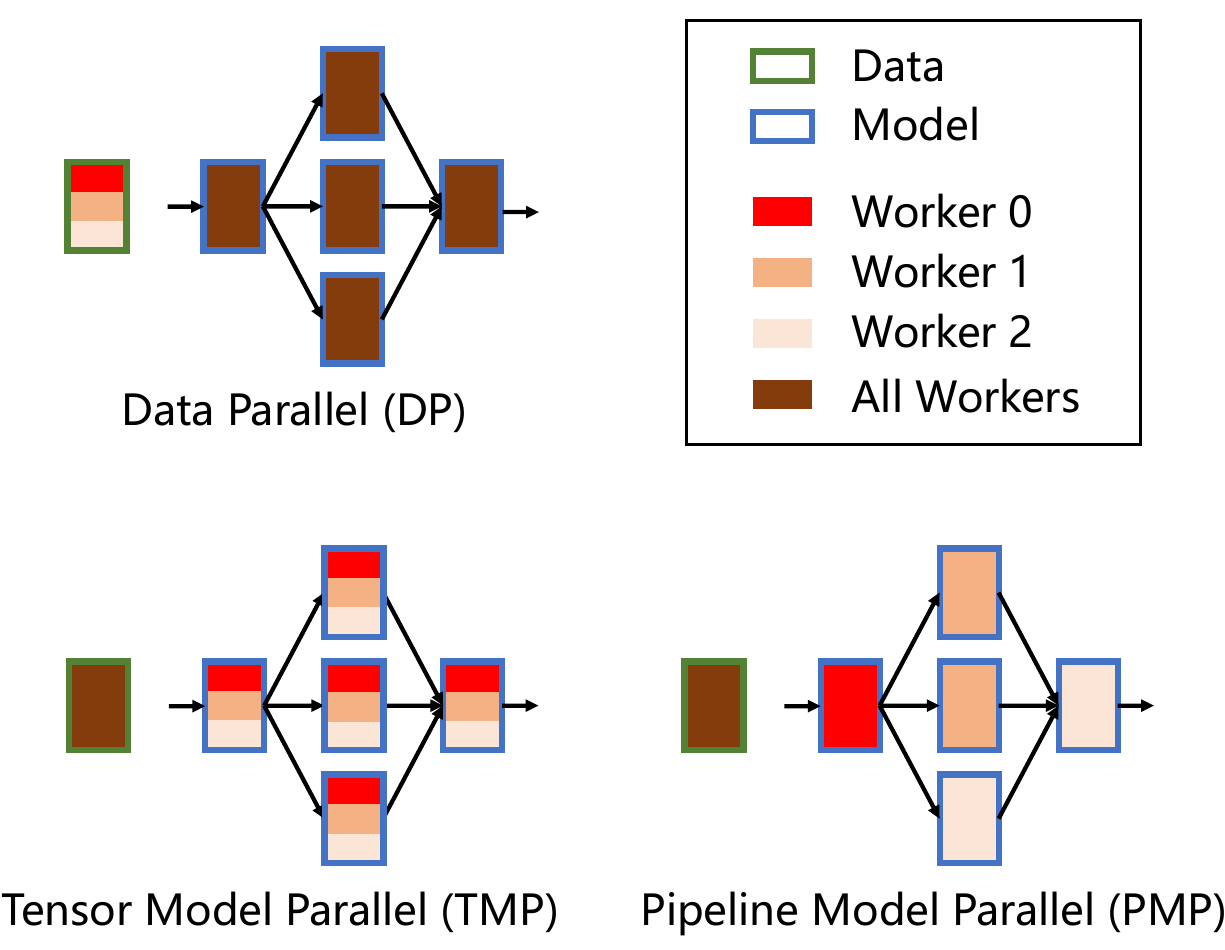}
	\caption{Common-used parallel strategies of neural network training.}
	\label{figArchDPMP}
\end{figure}

For running on an accelerator cluster, the AI tasks are usually executed under the data parallel (DP) \cite{NeurIPS12-DP} or model parallel (MP) \cite{NeurIPS12-MP} strategy, as shown in Figure \ref{figArchDPMP}.
The accelerator cluster can be seen as a graph where each node represents one accelerator (or some tightly linked accelerators), and each edge represents the data path between nodes.
Under the DP strategy, samples are distributed to nodes (i.e. different nodes hold different parts of samples), while weights are broadcast to all nodes (i.e. each node processes on a copy of all weights).
In contrast, under the MP strategy, weights are divided in the dimension of input/output channel (Tensor MP) or layer (Pipeline MP) and distributed to nodes, and each node processes on all samples.

Since the training of neural networks on multiple nodes usually require all-reduce operation (i.e, collecting partial results from all nodes and distributing the updated data to all nodes)\cite{nccl,patarasuk2009bandwidth,ying2018image,cho2019blueconnect}, a powerful and stable network with high bandwidth should be built to link all the accelerators.
Taking NVIDIA GPU cluster as an example, the nodes are usually linked with a Fat-tree network \cite{FatTree} ensuring that each node can access any other one in full bandwidth.
Based on InfiniBand Quantum-2 switches\cite{nvidiaQuantum2,lu2022survey,cisneros2022nvidia} or Ethernet data center switches of equivalent performance, employing a three-layer Fat-tree network architecture, a single cluster can support a maximum of 100,000 GPU cards.

Many works have pointed out that bandwidth is a major bottleneck for GPUs\cite{won2023astra,zhang2020network,pati2023computation}. 
These works conduct experiments on GPUs and Wafer-scale Computing systems, scale the die-to-die bandwidth to up to 1 TB/s and demonstrate that larger bandwidth can effectively reduce the exposed communication time, thereby reducing the overall execution time.
Currently, NVIDIA's NDR InfiniBand supports 100 GB/s interconnection between every two cards in different nodes 
(the receiving and sending bandwidths are added together)\cite{nvidiaQuantum2,nvidiaGrace}.
Furthermore, NVIDIA has proposed NVLink\cite{NVLink}to provide 900 GB/s bandwidth between any two cards in a same node.
This is already a quite high number, but we have found that further increasing the bandwidth can lead to even greater benefits.
Based on Megatron-LM\cite{shoeybi2019megatron,narayanan2021efficient},  
we built a performance model in the scenario of training GPT-3\cite{gpt3}on an GPU cluster with 16,384 H100 GPUs (each node has 256 GPUs).
We found that, if the NVLink bandwidth is increased from 900 GB/s to 2 TB/s, theoretical training time can be reduced by 36.9\%, resulting in a speedup of 58.6\%.
Therefore, it is highly valuable to continue increasing the bandwidth between accelerators.


\subsection{Advanced Packaging and Chiplet \label{secBkgAdvPack}}

\begin{figure}[htbp]
	\subfigure[]{
			\centering
			\includegraphics[width=.95\linewidth]{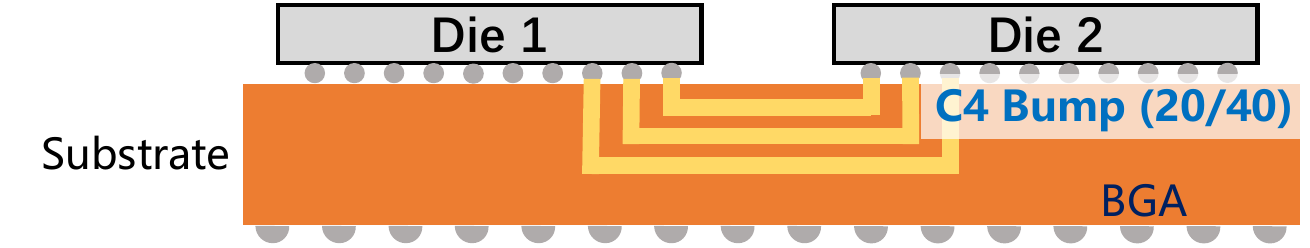}
			\label{figBkgAdvPack2a}
	}
	\subfigure[]{
			\centering
			\includegraphics[width=.95\linewidth]{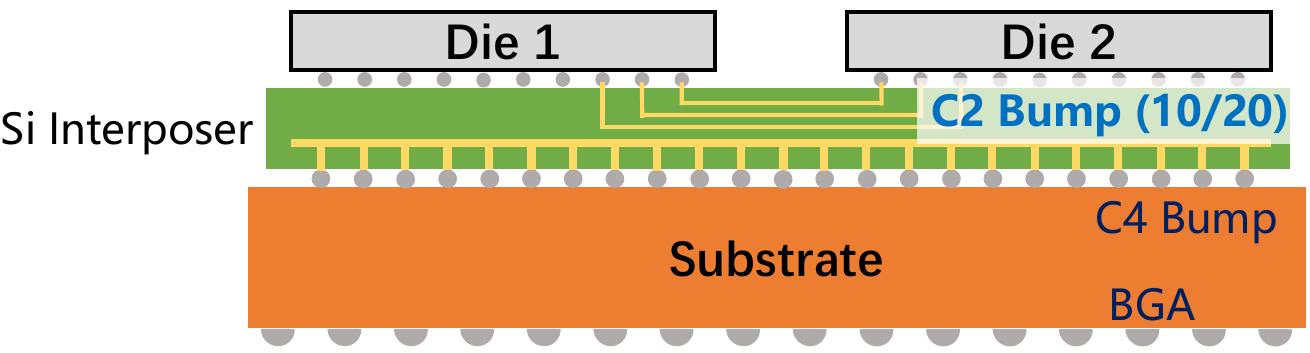}
			\label{figBkgAdvPack2b}
	}
	\subfigure[]{
			\centering
			\includegraphics[width=.95\linewidth]{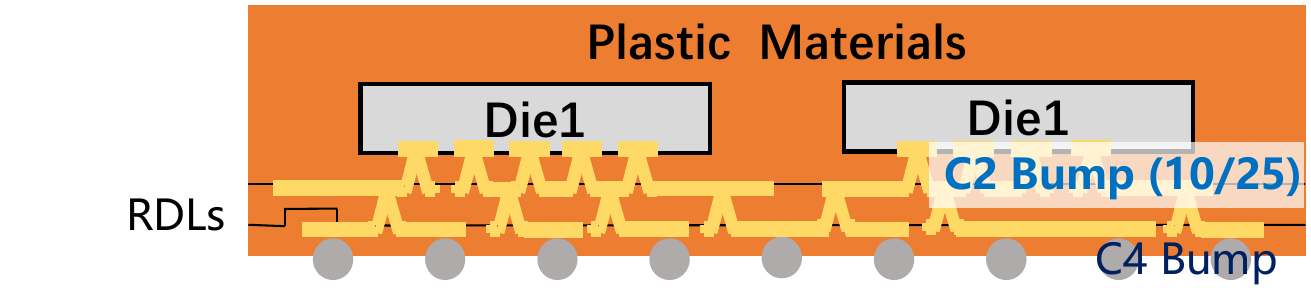}
			\label{figBkgAdvPack2c}
	}
	\caption{
Advanced packaging categories. (a) Substrate-based packaging. (b) Silicon
interposer-based packaging. (c) Redistribution layer (RDL)-based packaging. 
Blue font indicates the bump type and cd/pitch ($\mu m$) for die-to-die interconnection.
Adapted from \cite{ChipletSurvey1,kawano2021technology,Book21AdvPack}.}
	\label{figBkgAdvPack2}
\end{figure}

To increase the bandwidth between accelerators, a fundamental way is to reduce the physical distance between dies. 
This is also one of the primary goals of the advanced packaging technologies\cite{Book21AdvPack,lau2022recent,lee2020multi} that have garnered significant attention in recent years.
Unlike traditional accelerator architectures which package every single die into an individual device, 
advanced packaging architectures integrate multiple bare dies together (side by side or stacked vertically) and package them in a whole.
As shown in Figure \ref{figBkgAdvPack2}, the various advanced packaging technologies can be generally classified into three categories in terms of the carrier type: substrate-based, silicon-based, and redistribution layer (RDL)-based
packaging technologies.
The main advantages and disadvantages of each category are as follows:
\begin{itemize}
	\item Substrate-based packaging technology uses organic substrate materials to complete the wiring connections between dies with the etching process, which does not rely on the chip foundry process, so the related materials and production cost is low. However, the density of IO pins is low and the transmission capability per pin is affected by the crosstalk effect, so the bandwidth of die-to-die connections is limited.
    \item Silicon-based packaging technology implements the interconnection between dies by placing an extra silicon layer between the substrate and die.
    The connection between die and substrate is achieved with through-silicon vias (TSVs) and micro-bumps, which have smaller bump pitch and trace distance, so the IO density is improved and transmission delay and power consumption are reduced. 
    However, silicon interposer relies on the chip foundry process so the cost is much higher than organic substrate.
    To alleviate this, researchers proposed some variations of the original silicon interposer technology, such as silicon bridge technology \cite{ECTC18SiBridge} which only integrates small thin silicon layers on the substrate for inter-die interconnection, and silicon interconnect fabric (Si-IF) technology \cite{ECTC17SiIF} which removes the organic substrate and assembles the dies at small spacing using fine pitch interconnects.
    \item RDL-based packaging technology deposits metal and dielectric layers on the surface of the wafer to form a redistribution layer for carrying the metal wiring pattern and rearranging the IO ports.
    At present, fan-out style \cite{Fanout} is common-used, which rearranges the IO ports on the loose area outside the die, to shorten the circuit length for enhancing the signal quality.
    Compared to the silicon interposer-based packaging, RDL-based fan-out packaging has lower cost but less wiring resources.
\end{itemize}

The bare dies integrated by advanced packaging are also called as dielets or chiplets\cite{chiplet-dielet}, which introduces us to another well-known concept: chiplet technology\cite{ChipletSurvey1,ChipletSurvey2,ChipletSurvey3}.
Advanced packaging is a packaging technology aimed at integrating multiple functional components, 
while chiplet technology is a design methodology that divides an integrated circuit into multiple independent chips (chiplets), each containing specific functions or modules. 
The former serves as the infrastructure for the latter, and the latter is the driving force for the development of the former.

Originally, the principle of chiplet technology is based on the concept of modular design and integration. 
By dividing a complex integrated circuit into smaller chiplets each of which can be designed, manufactured, and tested independently, 
we can achieve easier customization, better reusability and scalability of chip designs, faster time-to-market, improved yield and cost efficiency.
Therefore, chiplet technology does not necessarily produce big chips. 
However, with the increasing demand for computing power, there have emerged many works that integrate large chiplets to break through the upper limits of single-chip computing power while keeping high internal bandwidth\cite{m1ultra,nvidiaGrace,samsungicube,2048chiplet,HCS22Dojo}.
Some of them even increase the total area to wafer level, for example, a Tesla Dojo Tile integrates twenty-five D1 dies with an area of 645 mm$^2$\cite{HCS22Dojo}, resulting in a total area exceeding 16,125 mm$^2$, which surpasses the area of the square inscribed in a 6-inch wafer (11,613 mm$^2$).
This kind of works can be referred to as wafer-scale chips.

\subsection{Wafer-scale Computing Systems \label{secBkgWSC}}

It should be noted that chiplet technology is just one way to achieve wafer-scale chips. 
There are other technological approaches, such as the field stitching\cite{ICWSI92Lithographic} used by Cerebras\cite{HCS22Cerebras}, that can also accomplish this. 
This computing paradigm, which scales the size of a single chip to the wafer level to achieve high computing power and large bandwidth benefits, is referred to as Wafer-scale Computing, regardless of the approach adopted.

More is different.
To extend the chip to the wafer level, there are many design issues need to be reconsidered. 
The major design consideration differences among various chip scales are shown in Table \ref{tabBkgDesignConsideration}.
While providing significant advantages on die-to-die bandwidth, integration density and programmability, wafer-scale chips also pose significant challenges on design space exploration, system implementation and tool chain development.

\begin{table*}
	\caption{Design Consideration Differences of Different Chip Scales \\
		 \label{tabBkgDesignConsideration}}
	\centering
	\scriptsize
	\begin{threeparttable}
		\begin{tabular}{c c c c}
			\hline
                & Monolithic Conventional Chip & GPU Cluster (One Node) & Wafer-scale Computing System (One Chip) \\
			\hline 
                Computing Power (FP16 Dense) & $\sim$1 POPS ** & $\sim$8 POPS *** & $\sim$9 POPS ****\\ 
                Monolithic Chip Area & $<$858 mm$^2$ & $<$858 mm$^2$ & $>$10,000 mm$^2$ \\ 
                Common-Used Memory System Pattern & on-Chip Shared SRAM + & on-Chip Shared SRAM + & on-Chip Distributed SRAM \\ 
                & off-Chip DRAM &  off-Chip DRAM &  \\
            \hline
                Inter-Die Interconnection & -- & Network Cable or NVLink & Advanced Packaging or \\
                 & & & Field Stitching \\
                Inter-Die Bandwidth  & -- & $\sim$0.9 TB/s & $\sim$8 TB/s \\
                Inter-Die Bandwidth Density  & -- & $<$10 TB/s/mm & $>$15 TB/s/mm \\
                Inter-Die Data Transfer Energy  & -- & $\sim$10 pJ/bit & $\sim$0.15 pJ/bit \\
            \hline 
                Fault Tolerance Design for Computation & Not Required & Not Required & Redundant Process Elements, \\
                & & & Bypass Routing, etc. \\ 
                Fault Tolerance Design for Data Transfer * & Not Required & Ethernet Protocol &  Redundant Physical Connections, \\ 
                & & & Redundant Routing, NoC Protocol, etc. \\ 
            \hline
                Power Delivery  & Conventional & Conventional & Edge or Vertical \\
                Clock Distribution & Conventional & Conventional & Edge or Vertical \\
                Cooling Solution & Conventional Air/Liquid & Conventional Air/Liquid & Conventional Air/Liquid or \\
                & & & Microfluidics \\
			\hline 
                Compilation  & Single-chip Mapping & Multi-chip Mapping & Multi-chip Mapping with finer grain and 2D-constraint \\
            \hline
		\end{tabular}
   		\begin{tablenotes}
			\footnotesize
            \item[] * \hspace{2em} Besides conventional ECC for DRAM.
            \item[] ** \hspace{1.5em} E.g., 1 NVIDIA H100 GPU \cite{Online22NVIDIAH100}: 989 TOPS.
            \item[] *** \hspace{1em} E.g., 1 DGX H100 = 8 NVIDIA H100 GPUs \cite{Online22NVIDIAH100}: 7.9 POPS.
            \item[] **** \hspace{0.5em} E.g., Tesla Dojo Tile \cite{HCS22Dojo}: 9.1 POPS. 
		\end{tablenotes}
	\end{threeparttable}
\end{table*}

\begin{figure*}
	\centering
	\includegraphics[width=.98\linewidth]{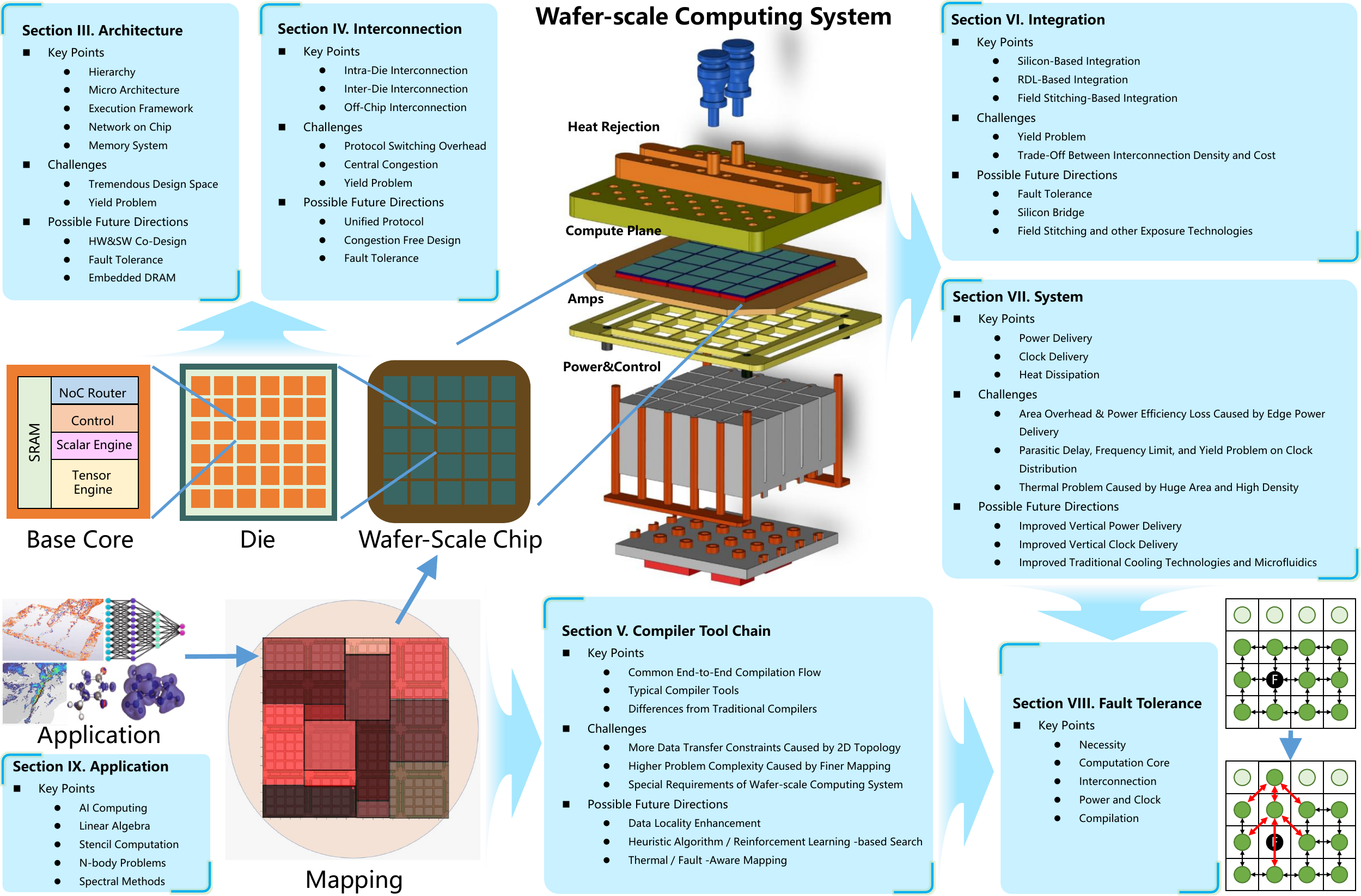}
	\caption{Overview of Wafer-scale Computing system and main contents of this paper.}
	\label{figBkgOverview}
\end{figure*}

Figure \ref{figBkgOverview} shows an overview of typical Wafer-scale Computing system, as well as the main contents of this paper.
The wafer-scale compute plane is the core component of the whole system.
A lot of effort needs to be paid into designing the architecture elements of it, including hierarchy, micro architecture, execution framework, NoC and on-chip memory system.
The interconnection interface and protocol design (including intra-die, inter-die and off-chip interconnections) are also extremely critical to the performance of wafer-scale chips.
To integrate the dies and interconnections together to form the physical wafer-scale chip, advanced packaging \cite{Book21AdvPack} or field stitching \cite{ICWSI92Lithographic} technologies are required.
Then, to drive the wafer-scale chip, specially designed power and clock delivery mechanisms are required, and the cooling modules is also necessary because of the high density of heat power.
At last, to run diverse applications on this Wafer-scale Computing system, a dedicated compiler tool chain for mapping the workloads onto the chip is indispensable.
In addition, fault tolerance designs relate to almost all aspects.
In the rest of this paper, we will discuss the key points, challenges and the possible future research directions from these aspects.

\section{Architecture \label{secArch}}

Like the traditional chips, the architecture design of wafer scale chips is also aimed at improving the performance, efficiency and flexibility.
But the difference is, designers should not only consider the trade-off within a single, conventionally sized die, but also the co-design of a tens of times larger system.
In this section, we will introduce the architecture design of existing typical wafer-scale works from the aspects of overall architecture, microarchitecture, dataflow, NoC, and memory system.

\subsection{Hierarchy \label{secArchOverall}}

\begin{figure*}
	\centering
	\includegraphics[width=.98\linewidth]{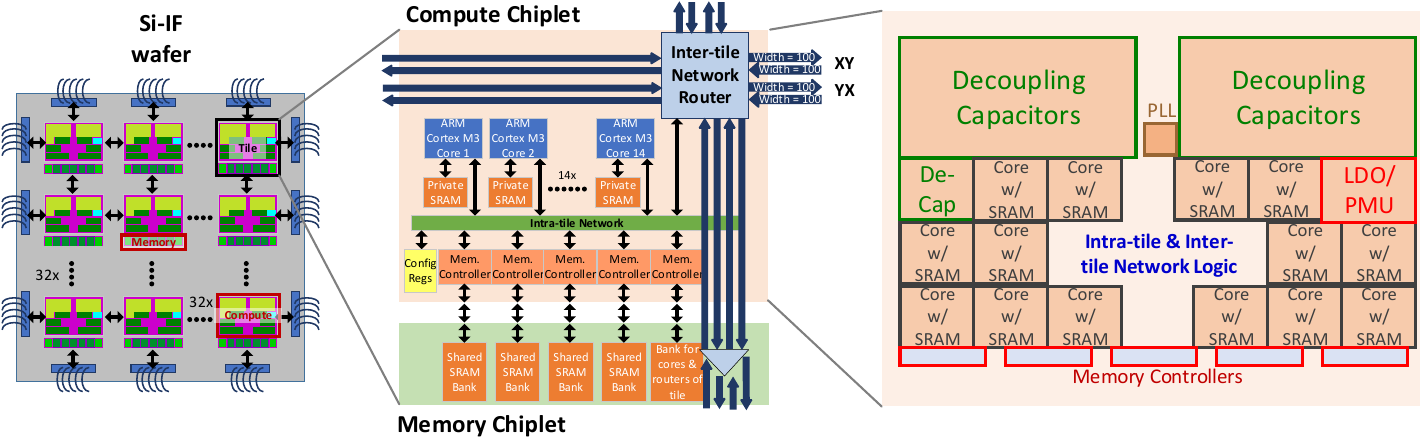}
	\caption{Overall architecture and microarchitecture of UCLA\&UIUC's work. Adapted from \cite{2048chiplet}.}
	\label{figArchUCLA}
\end{figure*}

\begin{figure*}
	\centering
	\includegraphics[width=.98\linewidth]{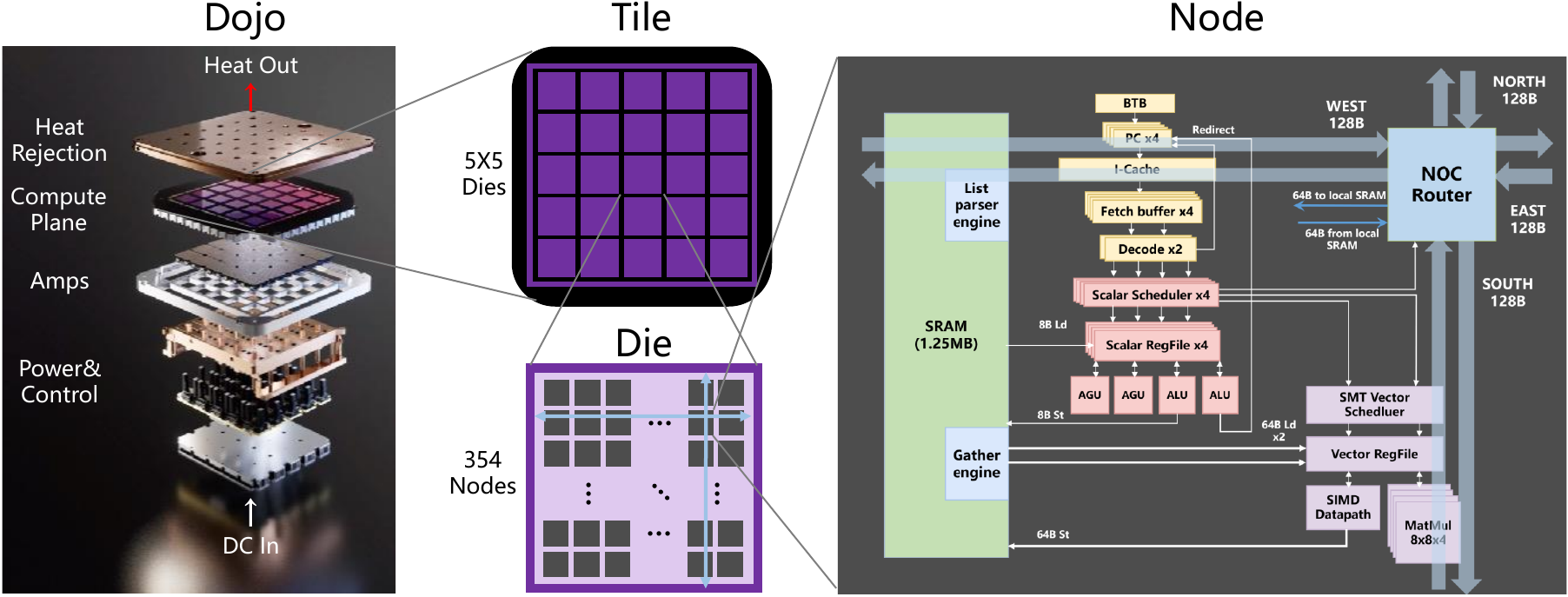}
	\caption{Overall architecture and microarchitecture of Tesla Dojo. Adapted from \cite{HCS22Dojo}.}
	\label{figArchDojo}
\end{figure*}

\begin{figure*}
	\centering
	\includegraphics[width=.98\linewidth]{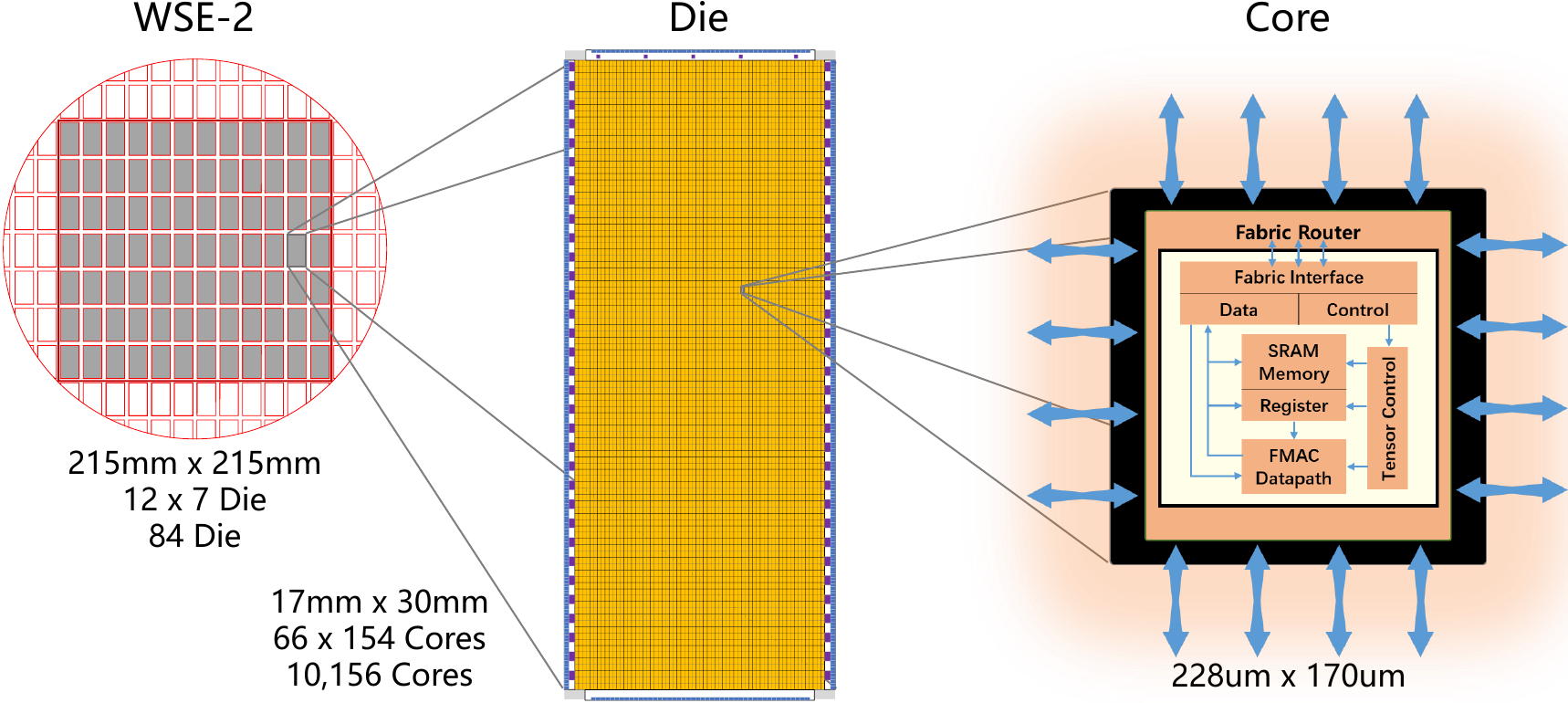}
	\caption{Overall architecture and microarchitecture of Cerebras CS-2. Adapted from \cite{HCS22Cerebras}.}
	\label{figArchCS2}
\end{figure*}

First let's review the overall architectures of three typical wafer-scale systems, including UCLA\&UIUC's work\cite{2048chiplet}, Tesla Dojo \cite{HCS22Dojo} and Cerebras CS-2 \cite{HCS22Cerebras}, as shown in Figure \ref{figArchUCLA}, Figure \ref{figArchDojo} and  Figure \ref{figArchCS2}.

UCLA\&UIUC's wafer-scale processor is comprised of 32$ \times$ 32 tiles, each tile is heterogeneously integrated by a compute chiplet and a memory chiplet.
The compute chiplet contains 14 ARM cores.
The total number of cores in the wafer-scale processor is 14,336.
Any core on any tile can directly access the globally shared memory across the entire wafer-scale system using the wafer-scale interconnect network.
Tesla's wafer-scale chip called Dojo training tile is comprised of 25 D1 dies and 40 I/O dies, and each D1 die contains 354 nodes.
DOJO nodes are full-fledged computers, and the total number of them in the Dojo training tile is 8,850.
Cerebras CS-2 is comprised of 12$\times$7 dies, and each die is comprised of 66$\times$154 cores.
Cerebras adopts more granular cores than UCLA\&UIUC and Tesla, the total number of cores in Cerebras CS-2 is as high as 853,104.

The characteristics of the three wafer-scale works are listed in Table \ref{tabArchOverallCharac}.
In summary, all the three wafer-scale works integrate a large number of cores to form an over 15,000 mm$^2$ wafer, aggressively use on-wafer distributed memories, and adopt a mesh/torus topology at the top level.
The computing powers of all the three wafers reach several POPS.




\begin{table*}
	\caption{Characteristics of typical wafer-scale systems\\
		 \label{tabArchOverallCharac}}
	\centering
	\scriptsize
	\begin{threeparttable}
		\begin{tabular}{c | c c | c c | c c}
			\hline
			& \multicolumn{2}{c|}{UCLA\&UIUC's work \cite{2048chiplet}} & \multicolumn{2}{c|}{Tesla Dojo \cite{HCS22Dojo}} & \multicolumn{2}{c}{Cerebras CS-2 \cite{HCS22Cerebras}}  \\ 	\hline
			Name of Wafer / Base Core & Array & ARM Core & Dojo Tile & Node & WSE-2 & Core \\ 
            Area & 15,100 mm$^2$ & $<$0.54 mm$^2$\tnote{*}  & $>$16,125 mm$^2$ & $<$1.82 mm$^2$\tnote{*}  & 46,255 mm$^2$ & 0.03876 mm$^2$ \\ 
			Comp. Power & 4.4 POPS & 307 GOPS\tnote{*} & 9.1 POPS & 1.02 TOPS\tnote{*} & 7.5 POPS & 8.8 GOPS\tnote{*} \\ 
			SRAM & 896+512 MB & 64 KB & 11 GB &1.25 MB & 40 GB & 48 KB \\
			SRAM BW & 6.144 TBps & --  & -- & 270-400 GBps &  -- & 23.5 GBps\tnote{*} \\
			Network BW & 9.83 TBps & --  & 9 TBps & -- & -- & 17.6 GBps\tnote{*} \\
   
            Power & -- & -- & 15 kW & $<$1.7 W\tnote{*} & 26 kW\tnote{**} & 30 mW \\
			\hline 
		\end{tabular}
   		\begin{tablenotes}
			\footnotesize
			\item[] *  \hspace{0.5em} Estimated through dividing the total amount by the number of base cores.
            \item[] ** \hspace{0.0000001em} Estimated through multiplying the value of base cores by the number of base cores.
		\end{tablenotes}
	\end{threeparttable}
\end{table*}

\subsection{Microarchitecture \label{secArchUArch}}


Now let's see the microarchitectures of aforementioned typical wafer-scale works, as shown in the right side of Figure \ref{figArchUCLA}, Figure \ref{figArchDojo} and Figure \ref{figArchCS2}.
The three wafer-scale works employ different strategies in base core designing.
UCLA\&UIUC's work \cite{2048chiplet} adopts a standard general-purpose processor (ARM Cortex M3) as the base core,
Cerebras CS-2 \cite{HCS22Cerebras} builds the base core around tensor acceleration,
and Tesla Dojo \cite{HCS22Dojo} designs a large full-fledged core which combines vector computing units and scalar general-purpose processing modules.
AI acceleration is one of the primary target applications of Tesla Dojo and Cerebras CS-2, so they both devote many resources on vector units for tensor acceleration.
The scalar general-purpose processing modules in Tesla Dojo are aimed at providing flexibility supports, such as branch jumping and sparse computing.
Cerebras CS-2 proposes a dataflow trigger mechanism to accomplish this.

The base cores of Tesla Dojo and Cerebras CS-2 are both homogeneously integrated to form a 2D mesh/torus grid, while the base cores in UCLA\&UIUC's work
are heterogeneously integrated with private SRAMs, shared SRAMs, and intra-tile networks, to form a tile, and then the tiles form a homogeneous 2D mesh grid.

\subsection{Execution Framework\label{secArchDataflow}}

\begin{figure}
	\centering
	\includegraphics[width=.98\linewidth]{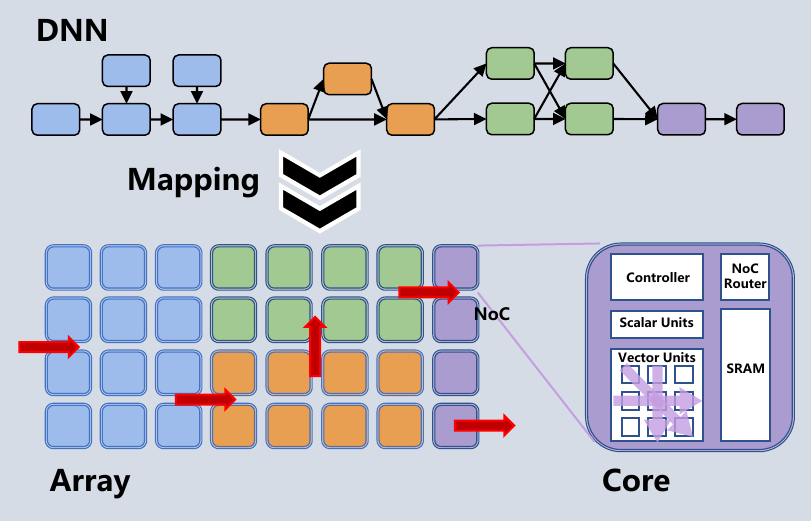}
	\caption{Typical execution framework of Wafer-Scale Computing systems.}
	\label{figArchUarchDataflow}
\end{figure}

To execute tasks, wafer-scale system can utilize multiple granularities of parallelisms, which is similar to GPU clusters \cite{pal2019architecting,Online21TeslaDojo,james2020ispd}.
As shown in Figure \ref{figArchUarchDataflow}, the wafer-scale chip can be divided into multiple partitions, each partition containing several cores works like a device in GPU cluster.
At macro level, the tasks (e.g. training of neural networks) are executed under the data parallel (DP) \cite{NeurIPS12-DP} or model parallel (MP) \cite{NeurIPS12-MP} strategy.
At micro level, each node processes the sub-task assigned to it like traditional accelerators do.
The vector units perform parallel computing on the partitioned tensors, and scalar units provide flexibility supports such as conditional branch jumping and sparse computing.

Compared with traditional accelerators, the differences include:
\begin{itemize}
	\item The much higher bandwidth of network on chip (NoC) and smaller gap between intra- and inter-die bandwidth broaden the design space of parallel.
	\item 2D-only interconnection imposes extra constraints on the design space. Specifically, long-distance data transfers on 2D mesh NoC would seriously waste bandwidth, so it is necessary to keep data locality, as emphasized by Tesla \cite{TeslaAIDay2021,HCS22Dojo}. 
    Usually, it is  better to adopt a dataflow-based execution framework which only passes data from one die to its neighbors if possible.
\end{itemize}

\subsection{Network on Chip (NoC) \label{secArchNoC}}

\begin{figure}
	\centering
	\includegraphics[width=.95\linewidth]{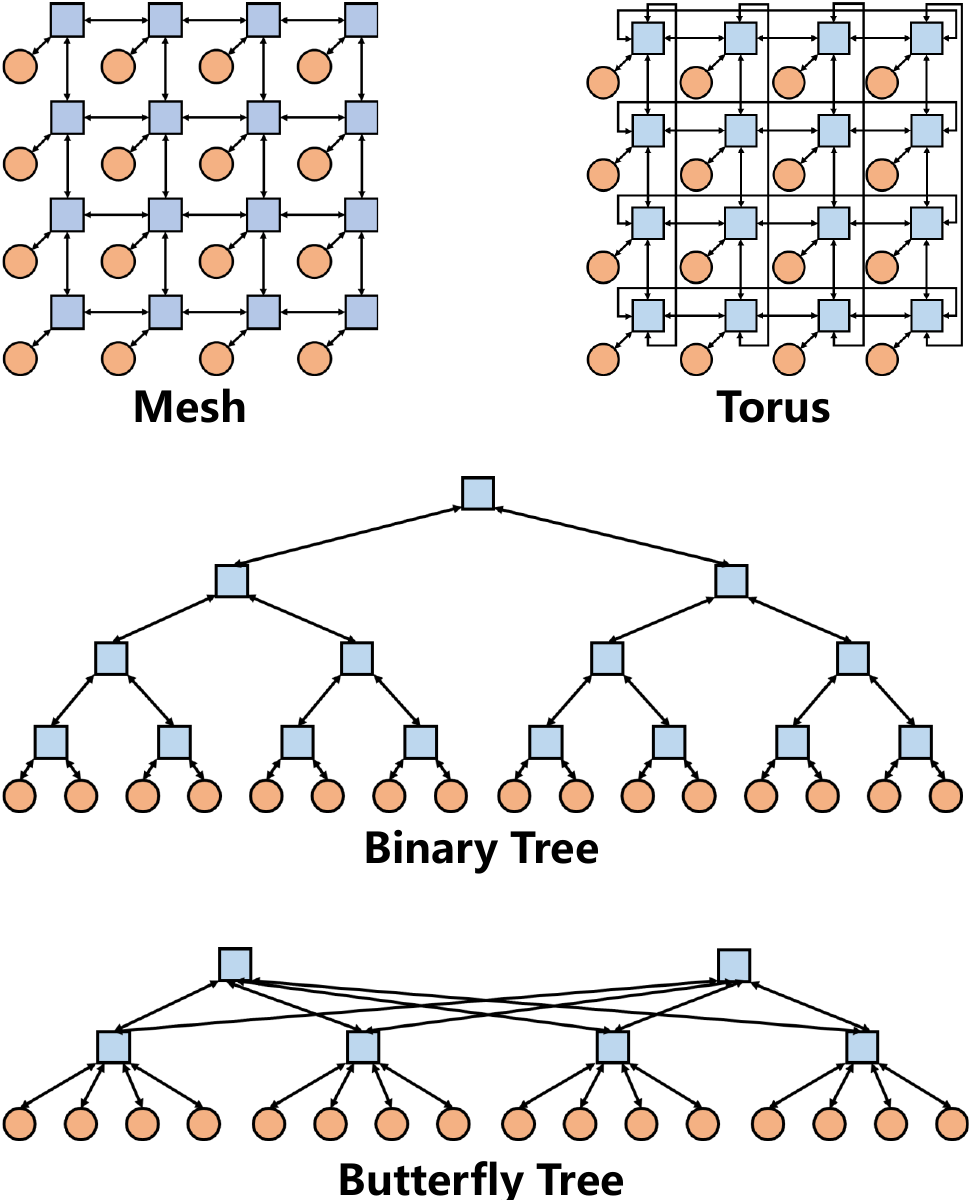}
	\caption{Common-Used NoC topologies. Adapted from \cite{cebry2020network}.}
	\label{figArchTopol}
\end{figure}

Topology is the first major design decision for a network on chip (NoC).
Common-used topologies include mesh, torus, binary tree, and butterfly tree \cite{cebry2020network}, as shown in figure \ref{figArchTopol}.
Topologies differ in the number of required routing units, the minimum and maximum number of connections or hops between any two compute units, as shown in Table \ref{tabArchNoC}.

\begin{table*}
	\caption{Characteristics of common-used NoC topologies \label{tabArchNoC}}
	\centering
	\scriptsize
	\begin{threeparttable}
		\begin{tabular}{c | c c c c}
			\hline
			Topology & \# Router Ports  & \# Routers & Min \# Hops & Max \# Hops \\ \hline
			Mesh & 5 & N & 3 & 	Height+Width+2\\
			Torus & 5 & N & 3 & Height/2+Width/2+2 \\
			Binary Tree & 3 & N-1 & 2 & 2log$_2$(N) \\
			Butterfly Tree & 6 & N/2-2 & 2 & 2[log$_2$(N)-2] \\
			\hline 
		\end{tabular}
	\end{threeparttable}
		\begin{tablenotes}
			\footnotesize
			\item[] Note: The number of node is N. The data come from thesis \cite{cebry2020network}.
		\end{tablenotes}
\end{table*}

\begin{figure}
	\centering
	\includegraphics[width=.95\linewidth]{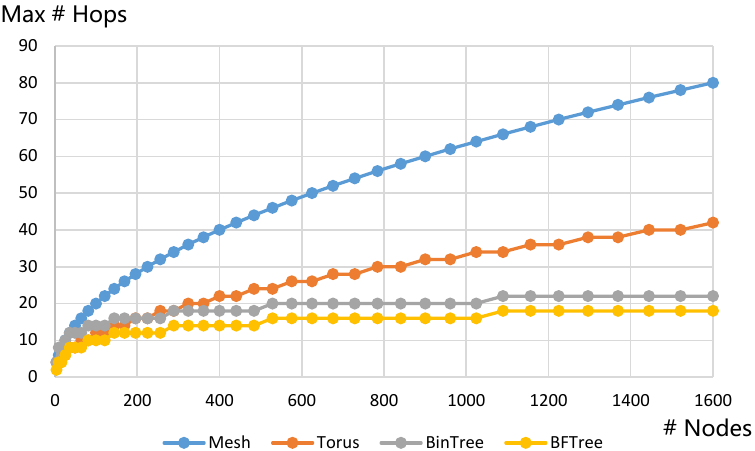}
	\caption{Maximum number of hops VS number of nodes.}
	\label{figArchNVH}
\end{figure}

Generally speaking, tree topologies tend to have smaller number of hops than mesh and torus, especially in large networks, as shown in Figure \ref{figArchNVH}.
However, mesh and torus topologies are more popular in existing wafer-scale systems.
The networks of UCLA\&UIUC's work\cite{2048chiplet} (inter-tile), Tesla Dojo \cite{HCS22Dojo} and Cerebras CS-2 \cite{HCS22Cerebras} all adopt a mesh or torus topology.
The main reasons why they prefer mesh or torus may include:
\begin{itemize}
	\item Less physical implementation difficulty: In mesh and torus topologies, the compute and routing unit pairs are arranged in a grid which makes it easier to integrate them in a wafer-scale chip and do scaling out.
	\item Better fault tolerance: When a router breaks down, it is easier to find alternate paths in mesh and torus topologies than tree topologies.
\end{itemize}

\subsection{Memory System\label{secArchMem}}

In Cerebras CS-2 \cite{HCS22Cerebras} and UCLA\&UIUC's work\cite{2048chiplet}, all memories are SRAMs.
Cerebras CS-2 fully distributes the SRAMs with cores.
Although UCLA\&UIUC's work adopts both private SRAMs for cores and shared SRAMs for global access, the shared SRAMs are also distributed with tiles.
Tesla Dojo \cite{HCS22Dojo} adopts both on-wafer SRAMs and off-wafer DRAMs (HBMs), and the SRAMs are also distributed with cores (nodes).

Generally speaking, existing wafer-scale system works \cite{2048chiplet,HCS22Dojo,HCS22Cerebras} tend to adopt distributed memory structures, here are the main reasons for it:
\begin{itemize}
	\item Attaching large DRAMs to each node of traditional accelerator clusters is a common practice, but it is difficult to integrate large DRAMs with each die on wafer-scale chips because of the limited area and the process differences between DRAM units and compute logics. As a result, the heavy storage burden for wafer-scale chips mainly falls on the distributed SRAMs. One might have noticed that Tesla also uses HBMs on the sides of Dojo tiles \cite{HCS22Dojo,HCS22DojoSystem}, but the cost of data transfer between HBMs and the central dies is high, so the locality of data access should be enhanced, which is reported by Dojo developers on AI day \cite{TeslaAIDay2021}.
	\item In a wafer-scale chip, centralized memory may cause seriously non-uniform access latency between the cores near and far away. Distributed memory structure and elaborated parallel strategy helps keep data locality and avoid most long-distance memory access, so that reduce the difficulty of scaling out.
	\item The bandwidth of NoC in wafer-scale chips is far more abundant than that in traditional accelerator clusters, which supports the use of distributed memory. As Cerebras points out in HotChips2022 \cite{HCS22Cerebras}, traditional centralized memory with low bandwidth requires high data reuse and caching to be efficient, while distributed memory with high bandwidth can realize full datapath performance without caching.
\end{itemize}

\subsection{Discussion \label{secArchDisc}}

\subsubsection{Design space exploration for wafer-scale chips}
It is a formidable task to design a wafer-scale chip and achieve the optimization. 
The design elements are closely related, and should be considered jointly.
Many of them are not included in traditional chip design and there is no experience that can be learned from.

The design space of wafer-scale systems can be composed of three main layers: workload, system and network \cite{won2023astra}.
In the workload layer, the superficial requirements (e.g. the target application tasks and the expected performance) should be transformed into high-level requirements (e.g. operator types, parallelization strategies, data communication and reuse patterns). 
In the system level, the microarchitecture and overall architecture (e.g.  dataflow design, compute units and memory organization) should be decided, to meet the requirements.
In the network level, network framework and implementation (e.g. hierarchical fabric design and topology, interconnect interfaces and protocols) should be decided to ensure the system runs as expected.
In addition, many extra variables are produced by the wafer-scale background, e.g. various implementations of die-to-die interconnection and fault tolerance mechanisms.

The design choices in the three levels interact with each other,
so separately deciding the design elements step by step may cause two problems:
1) Cumulative effect of minor losses \cite{pal2021scale}: the accumulation of small losses in the parts may cause a large loss on the whole system, resulting in 
unsatisfactory global solutions.
2) Ignorance of system bottleneck: the optimizer in one level cannot get exact information in other levels so that it does not know what the real system bottleneck is, and may make much useless work in the non-critical paths, leading to non-optimal choices.
However, exploring the whole design space of wafer-scale chips including all the three levels at once will suffer from a tremendous time complexity.

Although Cerebras and Tesla have showed most of the detailed parameters of their wafer-scale chips, they have not disclosed their design space exploration strategies.
The group from UCLA\&UIUC has presented their design space exploration framework \cite{pal2021scale}.
It is the first optimization framework for solving the multiple chiplet/system selection problem.
From the view of wafer-scale chip design, this framework has the advantage of supporting heterogeneous integration, but it does not cover important integration models such as Cerebras CS-2's reticle stitching technology \cite{ICWSI92Lithographic,MICRO21CBStory}.
Moreover, the reported results lack the cases of training large neural network models on real wafer-scale chips. 
How to abstract the requirements of such workloads, how to decompose the tasks and map to the cores, how to load data from off-wafer storages, how to design the dataflow and organize the data communication across multiple cores, these key points are unclear.
To fully exploit the advantages of wafer-scale integration, where the design choices in different levels are tightly coupled, cross-layer hardware and software co-design is required.
Therefore, a more comprehensive and mature methodology for wafer-scale chip design space exploration is yet to be developed.

\subsubsection{Design considerations related to the yield problem}
Compared with traditional chip manufacturing, building wafer-scale systems faces greater challenge of yield problem, so designers try to propose fault tolerance mechanisms to address it.
Cerebras proposes a fault tolerance mechanism with redundant cores and redundant fabric links \cite{HCS19Cerebras,lie2022processor}.
The redundant cores can replace defective cores and the extra links can reconnect fabric to restore logical 2D mesh.
Unlike UCLA\&UIUC's work and Tesla Dojo which use individual pre-tested chiplets (i.e., known-good-dies, KGDs) for assembling into a wafer-scale chip, Cerebras directly produces an integrated wafer-scale chip, so the yield challenge is more significant.
Therefore, Cerebras' cores are designed to be quite small (much smaller than the cores of UCLA\&UIUC's work and Tesla Dojo), in order to address yield at low cost.
UCLA\&UIUC's fault tolerance mechanism \cite{2048chiplet} mainly focus on die-to-die interconnection.
Two independent networks across the wafer are designed, one with X-Y dimension-ordered routing and the other with Y-X dimension ordered routing, to ensure the access between any two tiles.
There is still much room for existing fault tolerance mechanisms to be improved.
The fault can be caused by defective links, chiplets, cores, or partial logics in a core.
Different problems should be solved in different ways, and their interactions need to be considered.
A complete set of fault tolerance technologies for wafer-scale systems have yet to be developed.

Yield problem is also related to other design points.
In existing wafer-scale system works \cite{2048chiplet,HCS22Dojo,HCS22Cerebras},
the NoC router and compute part of each core are designed to be highly decoupled, this facilitates the the implementation of fault tolerance mechanism (and the utilization of distributed memories).
Besides, existing wafer-scale system designs \cite{2048chiplet,HCS22Dojo,HCS22Cerebras} only integrate SRAMs on-chip, which are low-density and expensive, because the wafer-scale integration technology has not been mature enough to solve the process differences between DRAMs and compute logics while maintaining the yield. This may be one research direction in the future.


\section{interconnection Interfaces and Protocols \label{secInterconnect}}

The interconnection interface and protocol design (including intra-die, inter-die and off-chip interconnections) are extremely critical to the performance of wafer-scale chips.
Compared with traditional chips, we should consider not only the unit transmission bandwidth, power consumption per bit, but also the requirements of advanced manufacturing process, packaging technology and system integration.

In this section, we will first review common-used interconnect interfaces and protocols which may be used in wafer-scale chips, and then discuss what characteristics of existing interfaces and protocols are especially important for building Wafer-scale Computing systems, or how they can be tailored to fit for Wafer-scale Computing systems.

\subsection{Intra-Die Interconnection}

The intra-die interconnection is responsible for communication among processing elements and memories within each die.
For wafer-scale chips, the optional styles of intra-die interconnection include bus, crossbar, ring, network-on-chip (NoC), and so on, which are similar to those for conventional chips.
The bus is the simplest to design but hard to scale with the number of processing elements,
while the NoC is the opposite \cite{MultiCoreSurvey}.
There have been some typical, common-used protocols for bus, such as the Advanced Microcontroller Bus Architecture (AMBA) Advanced High-Performance bus (AHB) and Advanced Extensible Interface (AXI), Wishbone Bus, Open Core Protocol (OCP) and CoreConnect Bus \cite{BusSurvey}.
In contrast, the protocol for NoC is more complex and usually needs to be specifically designed according to the application scenario.
Designers should make decisions on many design elements to balance the performance, cost and robustness, such as
number of destinations (unicast or multicast), routing decisions (centralized, source, distributed or multiphase), adaptability (deterministic or adaptive), progressiveness (progressive or backtracking), minimality (profitable or misrouting) and number of paths (complete or partial)
\cite{NoCSurvey}.

\subsection{Inter-Die Interconnection}

The inter-die interconnection interfaces and protocols are used for interconnection multiple dies in a same package. 
Serial interface and parallel interface are two options of die-to-die physical layer interface used for wafer-scale chips.

\subsubsection{Inter-Die Serial Interfaces}

Serial interfaces can be implemented at a long interconnection distance, so it is the mainstream in the field of wafer-scale chips.

\textbf{USR SerDes: }
SerDes \cite{stauffer2008high,ChipletSurvey1} contains a series of serial interfaces ranging from long reach to short reach, including USR (Ultra-Short Reach), which focuses on high-speed die-to-die connection at ultra-short distance (10 mm level) by 2.5D/3D heterogeneous integration. Because of its short interconnection distance, USR is able to provide $<$ 1 pJ/bit power consumption, nanosecond-level latency, and low error rate via advanced coding, multi-bit transmission and other technologies. However, due to its short transmission distance demand, implementation of USR in large-scale integration such as wafer-scale chips could be challenging.

\textbf{Apple UltraFusion: }
Apple’s M1 Ultra chip \cite{Micro22AppleM1} uses TSMC's 5nm process. It includes 20 CPU cores, 64 GPU cores, 32 neural network engine NPU cores, 128GB of memory, and 800GBps of memory bandwidth. It also solves the interconnection of very-large area Chiplets.
Especially, the UltraFusion interconnection technology design of die-to-die is extremely impressive. According to the current published papers and patents, UltraFusion should be an interconnection architecture based on TSMC’s CoWoS-S5 interconnection technology, using Silicon Interposer and Micro-Bump technology. UltraFusion also provides more than 10,000 die-to-die connection signal lines, with an ultra-high interconnection bandwidth of 2.5TB/s. Compared to other multi-chip module (MCM) packaging technologies, UltraFusion's interposer provides dense and short metal interconnects between logic dies or between logic dies and memory stacks. It has better inter-chip integrity, lower power consumption, and can run at higher clock frequencies.
In addition, UltraFusion effectively improves packaging yield and reduces costs through Die-Stitching technology. In UltraFusion, only the Known Good Die (KGD) is bonded, which avoids the problem that the failed chips in traditional Wafer on Wafer (WoW) or Chip on Wafer (CoW) are encapsulated, thereby increasing post-package yield and reducing overall average cost.

\textbf{AMD Infinity Fabric: }
Infinity Fabric (IF) \cite{beck2018zeppelin} is AMD's proprietary system interconnect, consisting of two systems, Infinity Scalable Data Fabric (SDF) and Infinity Scalable Control Fabric (SCF), which are responsible for data transmission and control respectively. SDF has 2 distinct systems on-die and off-die. The on-die SDF connects CPU cores, GPU cores, IMC and other parts within the chip, while the off-die SDF is responsible for interconnection of die-to-die on the package or multiple sockets.

\subsubsection{Inter-Die Parallel Interfaces}

Compared to serial interfaces, parallel interfaces can deliver higher bandwidth, lower latency and lower IO power consumption, which are attractive for the die-to-die interconnection in wafer-scale chips. However, the requirement of massive transmission wires and extremely short interconnection distance has made it hard to apply in practice so far. In the future, when the process can achieve relatively high wire density and close inter-die distances, parallel interfaces can be taken into consideration.

\textbf{Intel AIB/MDIO: }
Intel’s AIB/MDIO \cite{lau2019overview} provides a parallel interconnection standard for the physical layer. MDIO technology provides higher transmission efficiency, and its response speed and bandwidth density are more than twice that of AIB. AIB and MDIO are applied to advanced 2.5D/3D packaging technologies with short interconnection communication distance and high interconnection density, such as EMIB \cite{ECTC16EMIB} and Foveros \cite{ingerly2019foveros}.

\textbf{TSMC LIPINCON: }
TSMC’s LIPINCON \cite{lin20207} developed a high-performance parallel interconnection interface technology for the specific case of small chip memory interface. Through advanced packaging technology such as InFO and CoWoS, it can use simpler "clock forwarding" circuit, which greatly reduces the area overhead of the I/O circuit and its parasitic effects.

\textbf{OCP ODSA BOW: }
The BoW interface proposed by the OCP ODSA group focuses on solving the problem of parallel interconnections based on organic substrates \cite{drucker2020open}. There are three types of BoW, namely BoW-base, BoW-fast and BoW-turbo. BoW-Base is designed for transmission distances less than 10mm, using an unterminated unidirectional interface with a data rate of up to 4Gbps per line. The BoW-Fast uses a termination interface for cable lengths up to 50mm. The transmission rate per line is 16Gbps. Compared with BoW-Fast, Bow-Turbo uses two wires and supports bidirectional 16Gbps transmission bandwidth.

\textbf{NVLink-C2C: }
NVIDIA's NVLink-C2C \cite{NVLinkC2C} is a chip interconnection technology extended from NVLink \cite{NVLink} for custom silicon integration.
It is extensible from PCB-level integration, multi-chip modules (MCM), silicon interposer or wafer-level connections, allowing NVIDIA GPUs, DPUs, and CPUs to be coherently interconnected with custom silicon.
With advanced packaging, NVLink-C2C interconnection delivers up to 25X more energy efficiency and is 90X more area-efficiency than a PCIe Gen 5 PHY on NVIDIA chips.

\subsubsection{inter-die interconnection protocols}

There are a variety of different die-to-die interconnection protocols can be adapting to the physical layer interfaces. To choose a suitable die-to-die interconnection protocols for wafer-scale chips, 
the programmability, scalability and fault tolerance should be considered.

\textbf{Intel PCIe gen 4, 5, and 6: }
PCI express (PCI-e) \cite{fountain2005pci} is a general-purpose serial interconnection technology that is not only suitable for interconnecting GPUs, network adapters, and other additional devices, but also for die-to-die interconnection communication.
Since PCI-e Gen 3 was widely adopted, it has been developing rapidly. The transfer rate of PCI-e Gen 4 is 16 GT/s, and the theoretical throughput is 32 GB/s for 16x ports. PCI-e Gen 5, which was finalized in 2019, has twice the transfer rate of Gen 4 at 32GT/s, and the 16x port has a bandwidth of about 64GB/s. Currently, the latest PCI-e Gen 6 standard has now been released, which doubles the data transfer rate from 32GT/s to 64GT/s, providing approximately 128 GB /s of bandwidth for 16x ports. In addition, the PCI-e Gen 6 uses a new PAM4 encoding to replace the PCIe 5.0 NRZ, using a 1b/1b encoding scheme, which can encapsulate more data in a single channel in the same time. The PCIe 6.0 \cite{sharma2020pci} also introduces low-latency forward error correction (FEC) and related mechanisms to improve bandwidth efficiency and reliability.

\textbf{Intel UCIe 1.0: }
Intel, AMD, Arm, Qualcomm, TSMC and other industry giants jointly established the Chiplet Standard Alliance and officially launched the universal Chiplet high-speed interconnection standard "Universal Chiplet interconnection Express" (Universal Chiplet interconnection Express, UCIe), aiming to define an open, interoperable chiplet ecosystem standard \cite{sharma2022universal}. The UCIe 1.0 protocol uniformly uses Intel's mature PCIe and CXL (Compute Express Link) interconnection bus technologies. PCIe provides broad interoperability and flexibility, while CXL can be used for more advanced low-latency/high-throughput connections. The establishment of the UCIe standard maximizes the combination of the advantages of various fabs and technology companies, which is more conducive to the development of integrated circuit interconnection technology.

\subsection{Off-Chip Interconnection} 

Due to high-density integration, wafer-scale chips require external IO modules for inter-chip interconnection. PCIe is usually used for the external communication of the IO modules. For example, a wafer-scale chip is connected to the IO module through PCIe PHY in Dojo’s design. Besides PCIe, many other high-speed protocols can also apply to chip-to-chip interconnection, with different tradeoffs being taken into account.

\subsubsection{Off-Chip interconnection Interface}

~\\

\textbf{LR/MR/VSR SerDes: }
For chip interconnection based on PCB boards, it can be divided into long reach/medium reach/very short reach SerDes according to transmission distance between chips \cite{stauffer2008high,ChipletSurvey1}. Although this interconnection technology has the advantages of high reliability, low cost and easy integration, it is difficult to meet the high-performance interconnection communication requirements between wafer dies in terms of delay, power consumption and density.

\subsubsection{Off-Chip interconnection Protocol}

~\\

\textbf{Ethernet: }
EtherNet/IP \cite{spurgeon2000ethernet} is an industrial communication network managed and published by the ODVA specification, which is realized by combining the CIP protocol, TCP/IP, and Ethernet. Currently, the mainstream Ethernet protocols include 100/25 Gigabit Ethernet, 200 and 400 Gigabit Ethernet, and RoCE. The 100/25 Gigabit Ethernet provides the ability to obtain a large number of 25 Gbps ports from a single switch using 100 Gbps switches and fanout cables. 200 Gbps Ethernet uses four 50 Gbps lanes per port, whereas the original 400 Gbps standard would use eight 50 Gbps lanes, simply doubling the number of lanes in a port. RoCE is a protocol for realizing RDMA access/transmission through conventional Ethernet. Its basic idea is to encapsulate an InfiniBand transmission packet into a normal Ethernet packet at the link layer. It uses Direct Memory Access (DMA), where the network adapter can read and write directly from host memory bypassing the CPU core, reducing CPU load.

\textbf{NVIDA NVLink: }
NVIDIA's NVLink \cite{NVLink} is the first high-speed inter-GPU interconnection technology to improve interconnection communication between GPUs in multi-GPU systems. NVLink allows multiple GPUs into a unified memory space, allowing one GPU to work on the local memory of another GPU that supports atomic operations. NVIDA P100 is the first product equipped with NVLink 1.0, and its single GPU has a bandwidth of 160GB/s, which is equivalent to 5 times the bandwidth of PCIe 4. The NVIDA V100 is equipped with NVLink 2.0, which increases the GPU bandwidth to 300G/s, which is almost 10 times that of PCIe 4. Currently, NVIDIA A100 integrates the latest NVLink 3.0. A single NVIDIA A100 Tensor core GPU supports up to 12 NVLink 3.0 connections, with a total bandwidth of 600G/s per second, almost 10 times the bandwidth of PCIe 5. In addition, NVIDIA has developed a separate NVLink switch chip - NVSwitch. NVSwitch has multiple ports, which can integrate multiple NVLinks and can be used to interconnection multiple GPUs. The NVSwitch 2.0 enables high-speed communication between all GPUs simultaneously at 600 GB/s.

\textbf{ARM CCIX: }
Cache Coherent interconnection for Accelerators (CCIX) \cite{kim2022trends} is an open standard interconnection designed to provide a cache coherent interconnection between CPUs and accelerators. CCIX mainly includes protocol specification and transport specification, which supports cache coherence by extending the functions of transaction layer and protocol layer on the standard PCIe data link layer. CCIX usually uses two mechanisms to improve performance and reduce latency. One is to use cache coherence to automatically keep the caches of processors and accelerators coherent. The other is to increase the raw bandwidth of the CCIX link, taking advantage of the standard 16 GT/s transfer rate of PCIe 4, up to a maximum rate of 25 GT/s.

\textbf{OpenCAPI: }
To improve PCIe, IBM launched the Coherent Accelerator Processor Interface (CAPI) protocol \cite{stuecheli2018ibm}.
The CAPI1.0 multiplexes the physical layer, link layer and transaction layer of PCIe, and uses the Payload field of PCIe data packets to tunneling package cache coherence and CAPI control transactions. In later version of CAPI, it gradually evolved into OpenCAPI, which has its own physical, link, transaction layer and independent processing blocks.
OpenCAPI is an interface between processors and accelerators. It aims at being low-latency and high-bandwidth. It allows an accelerator (which could be an FPGA, ASICs, ...) to access the host memory coherently, using virtual addresses. An OpenCAPI device can also host its own memory, that can be accessed from the host.

\textbf{Intel CXL: }
Compute Express Link (CXL) \cite{van2019hoti} is an open standard for high-speed interconnects designed to address the interconnection gap between CPUs and devices and between CPUs and memory. Intel pointed out that PCIe is perfect for client machines without too many devices and too much memory, but when dealing with multiple devices that require bandwidth and huge shared memory pools, PCIe has huge delays and inefficient access mechanisms. 
CXL is designed to overcome these problems without giving up the best parts of PCIe - the simplicity and adaptability of its physical layer. CXL is based on the physical layer of PCI-e Gen 5 and uses the same physical and electrical interface. CXL provides protocols in three areas: CXL.io, CXL.cache, and CXL.memory. CXL.io is the simplest one, which is very similar to PCI-e, supporting the same features as PCI-e; CXL.cache is used to handle device access to local processor memory, for example, to allow network drives to access their own caches; CXL.memory is used to provide processor access to attached device memory (non-native memory).

\subsection{Discussion \label{secInterconnectDisc}}

\subsubsection{Uniformity}

Traditional HPC systems employ multiple existing protocols that match the bandwidth at different hardware scales to access varying levels of memory, aiming to attain data communication and scalability across hardware dimensions. However, in emerging wafer-scale chips where intra- and inter-die bandwidths converge, the inherent rate mismatch across protocols and complex protocol hierarchy mandating replicate data motion across memory and protocol packets will incur significant communication performance penalties. To avoid these pitfalls and fully harness the bandwidth of wafer-scale chips, adopting a uniform cross-scale communication protocol offers an effective approach. 
That is what Tesla does -- proposing Tesla Transport Protocol (TTP) to cover the interconnections of intra-D1-die, inter-D1-die, and inter-Dojo-tile\cite{HCS22DojoSystem,HCS22Dojo}.

\subsubsection{Compensating for 2D NoC}

Central congestion is a key challenge for designing the interconnection protocol of wafer-scale chips.
Specifically, the interconnection of data centers can be physically implemented in the 3D space, 
which means more free topologies (e.g., FatTree) can be adopted and the central congestion problem can be naturally avoided.
In contrast, the NoCs for wafer-scale chips are usually limited to a 2D topology, which results in the closer to the center the larger traffic there is, 
so the solutions for central congestion problem should be considered when designing the protocols.
To this end, Tesla proposes Z-Plane Topology and the corresponding protocol named "Tesla Transport Protocol over Ethernet(TTPoE)" 
to enable "shortcuts" across large network hops \cite{HCS22DojoSystem,HCS22Dojo}.

\subsubsection{Fault Tolerance Support}
Fault tolerance is another core issue for wafer-scale chips.
Since physical failures of dies and interconnects may appear during production and use,
the interfaces and protocols for wafer-scale chips should be specially designed to support the fault tolerance functions,
e.g., switching the reserved datapath when the active one is broken, routing bypass the faulty dies.

\section{Compiler Tool Chain \label{secTool}}

Like the traditional platforms aimed at large-scale Deep Learning acceleration, Wafer-scale Computing systems also need a compiler toolchain to meet the user requirements of extreme computing efficiency and ease of use.
The former requirement means that the application workloads should be perfectly optimized, scheduled and mapped, and the hardware resources are fully utilized.
The latter requirement means that the compile procedure should be highly automatic, flexible, end-to-end, without too much manual work.
These two requirements are fairly contradictory and researchers have paid much effort on them for traditional Deep Learning acceleration platforms \cite{compilerSurvey2019, compilerSurvey2020,compilerSurvey2023}. 
Compared with the compiler tool chain of traditional platforms, the one of Wafer-scale Computing systems can be generally similar in overall framework, but significantly different in detailed design considerations and strategies. 

In this section, we will first give an overview of the common end-to-end compilation flow for driving large-scale neural network acceleration platforms,
and then review typical compiler tools for traditional platforms and Wafer-scale Computing platforms.
At last we will discuss what make the compilation of Wafer-scale Computing different from traditional one.

\subsection{Common End-to-End Compilation Flow of Large-Scale Deep Learning Acceleration}

\begin{figure*}[htbp]
    \centering
	\hspace{2mm}
	\includegraphics[width=0.8\linewidth]{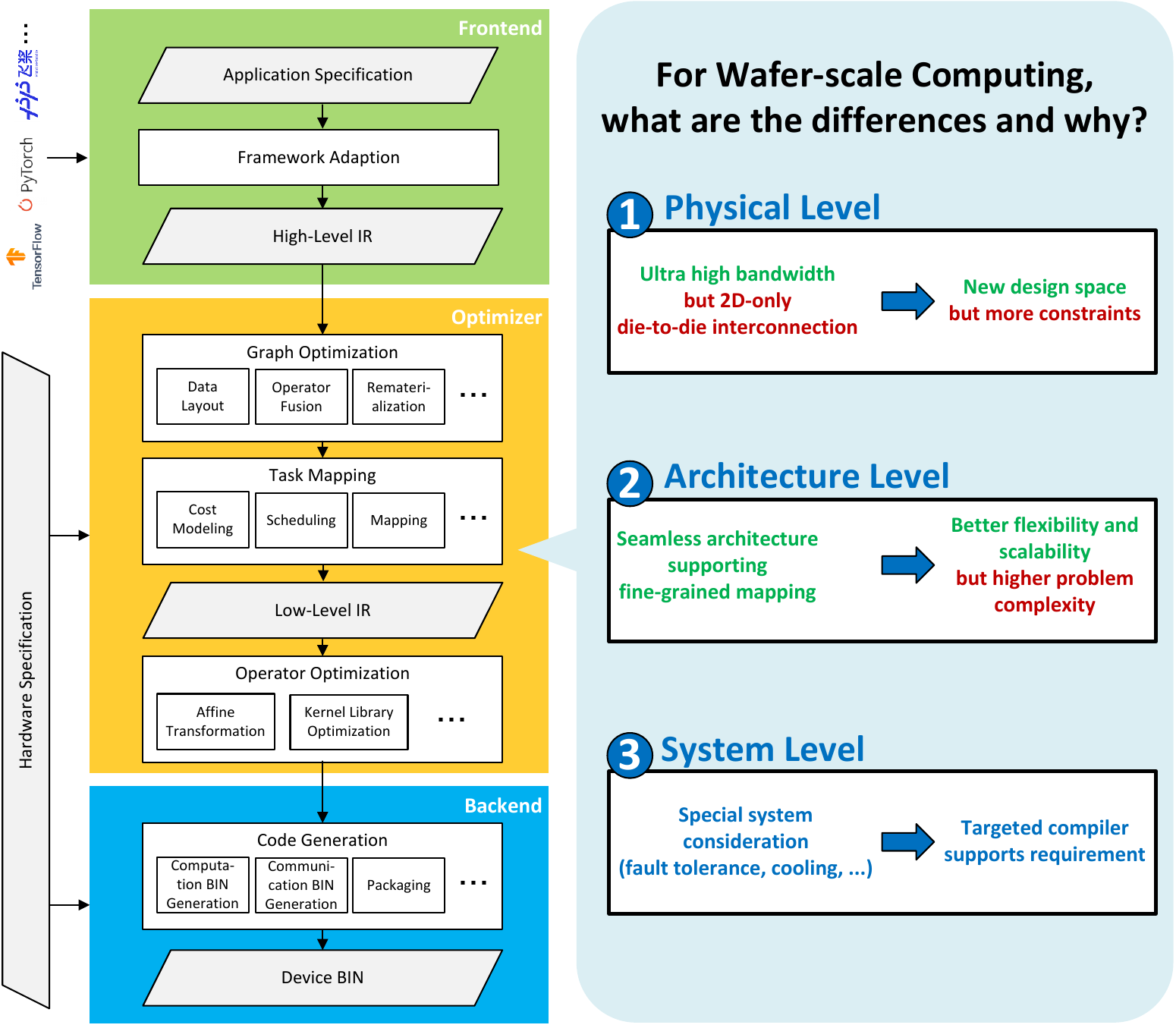}
	\caption{Overview of end-to-end compilation flow.
 }  
	\label{fig:CompilerFlow}
\end{figure*}

The common end-to-end compilation flow of Deep Learning acceleration is shown in Figure \ref{fig:CompilerFlow}.
First, the compiler transforms the inputted application specification into high-level intermediate representation (IR).
Since there are various deep learning frameworks (such as TensorFlow \cite{tensorflow}, Pytorch\cite{pytorch}, PaddlePaddle \cite{paddlepaddle}, and so on), the compiler usually needs to have the ability of framework adaption, i.e., be flexible enough for different forms of inputs.
The high-level IR typically comes in the form of directed acyclic graphs to represent the operations as nodes and the data dependencies as edges \cite{compilerSurvey2020}.
Then, the compiler performs graph optimizations on the high-level IR, such as operator fusion, data layout transformation and rematerialization \cite{chen2018tvm,cyphers2018intel,kirisame2020dynamic}.
After that, if the target hardware platform is an accelerator cluster with multiple devices, or a single device with multiple partitions, task mapping is performed, i.e., the task is divided into multiple sub-tasks, scheduled and mapped to the devices or partitions. Parallel and pipelining strategies are usually used to increase the hardware utilization and execution performance \cite{NeurIPS12-DP, NeurIPS12-MP}.
Usually, a high-level cost model is built to help decide which mapping solution to use.
After these stages, low-level IR is generated, 
which describes the operation in a more fine-grained representation than that in high-level IR, reflects the hardware characteristics and represents the hardware-specifc optimizations.
Based on it, operator optimization (e.g. affine transformation and kernel library optimization) is performed to tune the computation and memory access for each operator.
After completing the optimizations, the compile generates machine code from the optimized IR and for driving the hardware.

The whole compiler flow can be divided into three parts: frontend, optimizer(middle-end) and backend.
We are chiefly concerned with the optimizer in this section.
Generally speaking, graph optimization belongs to high-level optimization which takes multiple operators into consideration at a time, while operator optimization belongs to low-level optimization which focuses on how to execute a single operator more efficiently.
Graph optimization can be also regarded as part of the frontend, and operator optimization can be also regarded as part of the backend.

\subsection{Typical Compiler Tools for Large-Scale Deep Learning Acceleration}


In 2022, Zheng \emph{et al.} proposed an complier named \texttt{Alpa}, which can automatically generate model-mapping plans leveraging hierarchically optimizing the intra-operator and inter-operator parallelism \cite{zheng2022alpa}. 
The key insight is based on the observations that different parallelism schemes require diverse bandwidth for communication, while typical computing clusters exhibit corresponding structures that closely located cores communicate with sufficient bandwidth, while distant cores own limited bandwidth. 

\begin{figure}[htbp]
	\hspace{2mm}
	\includegraphics[width=0.93\linewidth]{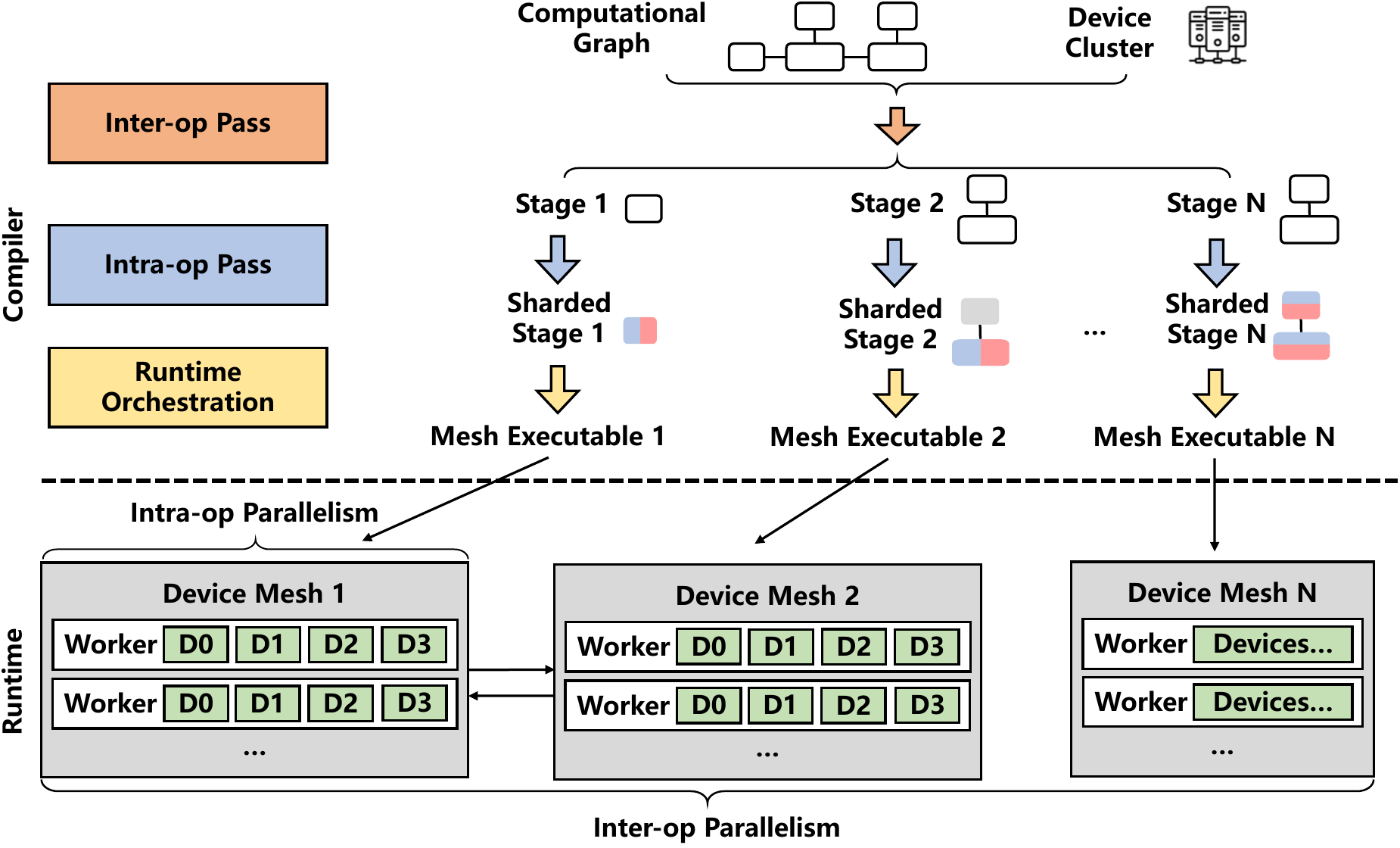}
	\caption{The complier flow and runtime architecture of the \texttt{Alpa}. Adapted from \cite{zheng2022alpa}.}
	\label{fig:Alpa1}
\end{figure}

Leveraging the asymmetric property of a computing cluster, the \texttt{Alpa} maps the intra-operator parallelism into the cores with high communication bandwidth, while allocating the inter-operator parallelism to those with limited bandwidth. Additionally, the \texttt{Alpa} utilizes a hierarchical architecture to express the plan in each parallelization category, which is depicted in Fig.~\ref{fig:Alpa1}. It is observed that given the computational graph and hardware configuration, the inter-operator compilation process separates the graph into a number of stages, while dividing the clusters into several device meshes. Specifically, the inter-operator process employs the dynamic programming (DP) algorithm to allocate stages to corresponding meshes and activate the intra-operator process on each stage-mesh pair, in order to obtain the execution cost of this assignment. Next, the intra-operator process optimizes the execution efficiency about the stage running on its assigned device mesh, by minimizing the corresponding execution cost utilizing integer linear programming algorithm, then return the cost report to the inter-operator process. Through calling the intra-operator process for each stage-mesh pair repeatedly, the inter-operator process employs the DP to minimize the execution latency of inter-operator parallelism while acquiring the optimal partition scheme for stages and meshes. 

In 2021, Tan \emph{et al.} presented an automatic tool named \texttt{NN-Baton}, which focuses on the chiplet granularity exploration and workload orchestration on different computation levels~\cite{tan2021nn}. 
The high-level flow of the \texttt{NN-Baton} consists of three main components: hardware design space exploration, mapping analysis engine and cost evaluation module. The inputs include resource constraints, NN model descriptions and hardware descriptions, where the resource constraints are primarily composed of the number of MAC units and the area budgets. Furthermore, the model description is parsed from the Pytorch model exerting the \texttt{torch.hit} function. 




The ISPD 2020 competition delivered a unique challenge targeting at allocating neural network workloads to the Cerebras CS-1 WSE~\cite{james2020ispd}. Since a vital feature of WSE lies in its sufficiently large capability to run every layer of an NN simultaneously, how to assign the workload to achieve a substantial increase in hardware efficiency is a critical question worths to be solved.

\begin{figure}[htbp]
	\hspace{1mm}
	\includegraphics[width=0.95\linewidth]{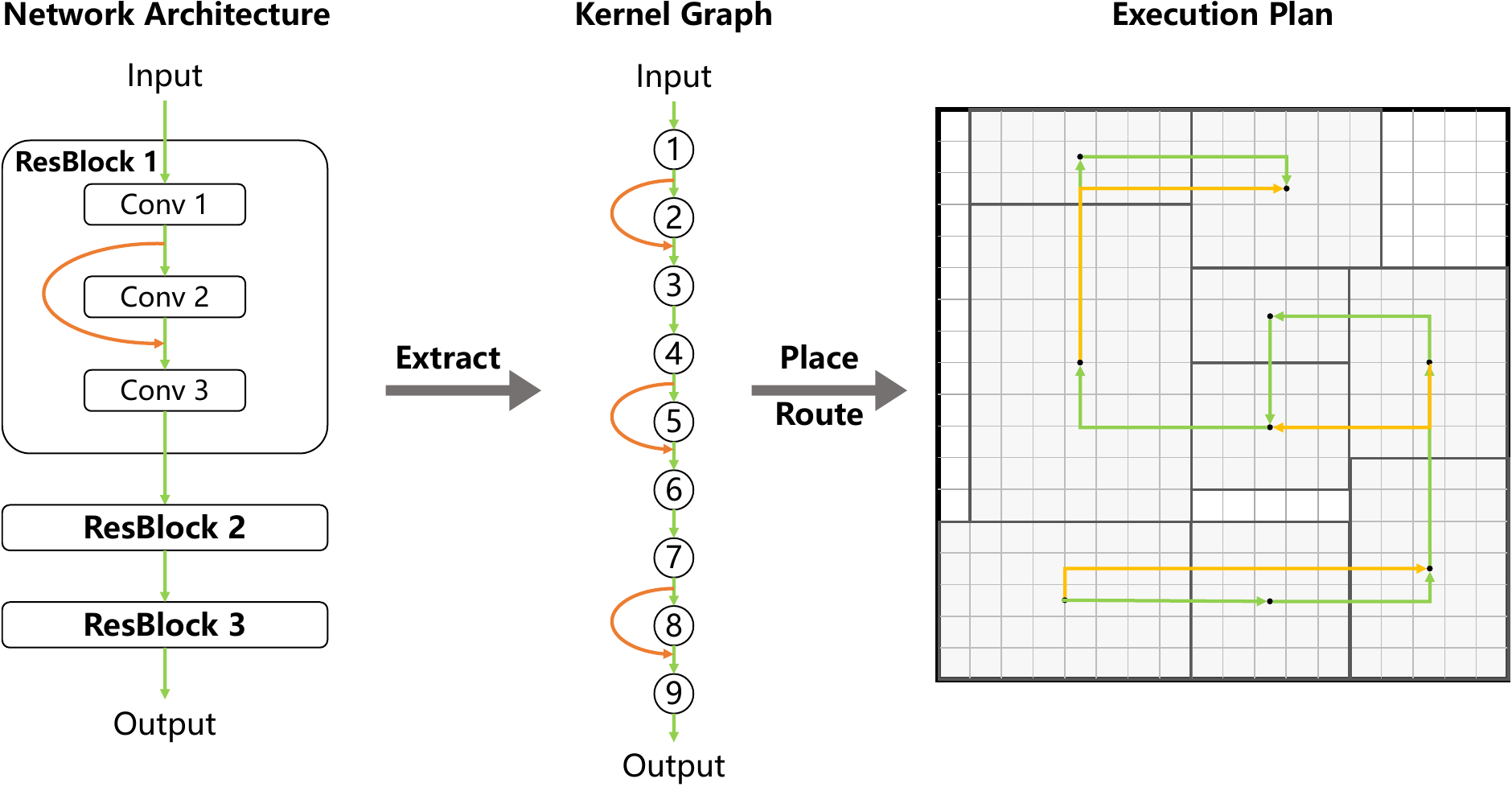}
	\caption{The overview of WSE compilation flow. Adapted from \cite{jiang2020cu}.}
	\label{fig:CUPOker1}
\end{figure}

Fig.~\ref{fig:CUPOker1} illustrates the corresponding compilation stage of WSE. As can be seen, the input NN models that usually are represented by ML frameworks will be converted into a graph representation exerting a set of predetermined kernels provided in the given kernel library. Specifically, the layers in the NN are mapped to kernels that each performs an individual computational tasks. For example, a kernel could compute a ``$3\times 3$ convolution", while the next kernel may execute a "$1024\to32$" fully connected layer. 

Next, the \emph{mapping algorithm} is employed to place and route the kernels of an NN on the computation fabric with specific objectives and constraints. Specifically, the solutions should satisfy the following constraints~\cite{james2020ispd}:
\begin{itemize}
	\item All kernels must fit the fabric area ($633 \times 633$ tiles).
	\item Kernels may not overlap.
	\item No kernel's memory exceeds the tile's memory limit.
\end{itemize}

While keeping the runtime of placement satisfies a specific time threshold, the quality of a feasible solution is evaluated by a weighted summation of the following objectives~\cite{james2020ispd}:
\begin{itemize}
	\item The maximum execution time (MET) among all kernels.
	\item The overall $\mathcal{L}_1$ distance of all connected kernels.
	\item The total adapter overhead of all connected kernels.
\end{itemize}

Since the complied model on WSE is executed in a pipeline fashion, the kernel with slowest throughput will limit the overall performance of the system. Additionally, $\mathcal{L}_1$ distance provides a simplified evaluation for the routing cost, and the adapter overhead reflects the cost required to unify the I/O protocols among kernels in practical systems.

Formally, a kernel represents a parametric program that performs specific tensor operations. Specifically, it is composed of a set of \textbf{formal arguments} to specify the shapes of the tensor operations to be executed, and a set of \textbf{execution arguments} to define how the operations are parallelized across the tiles~\cite{james2020ispd}. For given kernels, their formal arguments are specified by the input neural network specification and retain unchanged during the compilation procedure, while the execution arguments are configurable, whose values are variables to be optimized by the mapping algorithm.
Taking the convolution kernel as an example, it consists of eight formal arguments (H, W, R, S, C, K, T, U)  and four execution arguments $(h, w, c, k)$. Specifically, the (H, W) specify the size of input image in $2$-dimensions. (R, S) represents the size of convolution core in $2$-dimensions. (C, K) denotes the number of input and output features. (T, U) refer to the horizontal and vertical striding of the convolution operation. Moreover, $(h, w, c, k)$ express the unrolling of computations that can be performed in parallel. 

Jiang~\emph{et al.}~\cite{jiang2020cu} proposed a high-performance engine named \texttt{CU.POKer}, to fulfill the neural network workload placement tasks on WSE. 
First, the proposed engine adopts a \emph{binary search} and a \emph{neighbor-range search} to find a MET $T$. Then, for each given targeted $T$, kernel candidates with optimal shapes are generated. Next, a \emph{data-path-aware placement algorithm} is employed to place the generated kernels aiming at minimizing the total wirelength while guaranteeing no kernel's execution time exceeds the $T$. Considering the recursion employed in this step may result in computational exploration, some \emph{pruning} techniques are utilized. For example, a threshold $t_{\rm row}$ is set to adjust the maximum number of rows, which is initialized to one. If and only if no legal solution can be found under current threshold $t_{\rm row}$, the placement process will be invoked again with the threshold $t_{\rm row}+1$. Finally, a post refinement is executed to lower the overall adapter cost of the current solution.

\begin{figure}[htbp]
	\subfigure[Low wirelengths in each kernel with high inter-kernel wirelengths.]{
		\begin{minipage}{.46\linewidth}
			\centering
			\includegraphics[height=3.05cm]{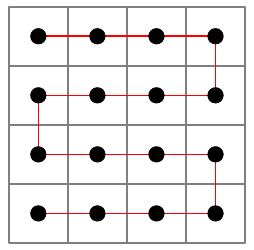}
			\label{fig:NewYork1}\vspace{1mm}
		\end{minipage}
	}\hspace{0mm}
	\subfigure[High wirelengths in each kernel with low inter-kernel wirelengths.]{
		\begin{minipage}{.46\linewidth}
			\centering
			\includegraphics[height=3.02cm]{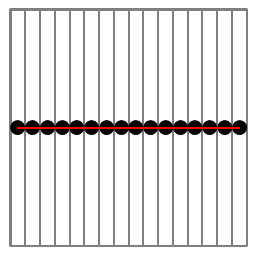}
			\label{fig:NewYork2}\vspace{1mm}
		\end{minipage}
	}\vspace{-1mm}
	\caption{The illustration for the shortcoming about cost functions in the ISPD 2020 contest. Adapted from \cite{ozdemir2022kernel}.}
	\label{fig:NewYork}
\end{figure}

However, the ISPD 2020 contest lacks consideration of internal wiring when improving performance. Specifically, the wirelength is the sum of $\mathcal{L}_1$ distance between the centers of all connected kernels. Accordingly, there is no penalty for kernel implementations that span the entire height of the wafer. The drawback with cost function in ISPD metric is depicted in the Fig.~\ref{fig:NewYork}. As can be seen, compared to Fig.~\ref{fig:NewYork2}, better internal wiring is acquired in Fig.~\ref{fig:NewYork1}, but the center-to-center inter-kernel distance metric penalizes Fig.~\ref{fig:NewYork1} severely. Moreover, considering the \texttt{CU.POker} utilizes a threshold to control the number of rows, more practical modeling would push the \texttt{CU.POKer} into the computational explosion.

To this end, an improved work for \texttt{CU.POker} is presented in~\cite{ozdemir2022kernel}. The improved  approach utilizes an initial topological structure similar to \texttt{CU.POKer}, but employs \emph{dynamic programming} to execute kernel selection and mapping. 
The corresponding experiments on the benchmarks in ISPD 2020 contest suite reveal that compared to the \texttt{CU.POKer}, the improved approach~\cite{ozdemir2022kernel} exhibits smaller execution time for all but two cases. Although the \texttt{CU.POKer} performs better on wirelengths, the authors think it is not aligned with realistic system performance considerations.

Additionally, Li \emph{et al.} present another mapping engine named \texttt{GigaPlacer} \cite{li2021placement}, which integrates binary search and dynamic programming algorithms to reduce the compilation time.
Peng \emph{et al.} combine graph theory and combinatorial optimization technology to devise a fast floorplanning approach ~\cite{peng2021resource}.

Although the ISPD contest successfully provides an available cost model for the wafer-scale chip, there are still some limitations to be further considered. For example, since diverse data access modes lead to different computing behaviors, the relationship among computation latency, data access latency, data transfer latency should be considered more holistically. Further, the existing considerations only lie in the inter-operator optimization. However, the intra-operator optimization needs to be further explored. Specifically, for a PE array with given shape, how to orchestrate the dataflow can minimize the cost of data transfer is a huge design space and need to be considered in more details. 

\subsection{The Key Differences of Wafer-Scale Computing Compilers from Traditional Ones \label{secToolDisc}}

Wafer-scale Computing relies heavily on software-hardware co-design, so the platform characteristics decide what are different for Wafer-scale Computing compilers.

\subsubsection{Physical Level}
As integrated tightly by advanced packaging, wafer-scale chips have much larger die-to-die bandwidth than traditional accelerator clusters (e.g., the die-to-die bandwidth of Tesla Dojo D1 die's every edge is 2 TB/s while the NVLink bandwidth of NVIDIA H100 is only 900 GB/s \cite{HCS22Dojo,Online22NVIDIAH100}), which breaks the previous memory wall and opens up new design spaces for exploring more aggressive application solutions that can fully utilize the computing power available. But on the other hand, wafer-scale chips limit the NoC topologies within 2D space (usually 2D mesh or 2D torus), which is the cost to have better die-to-die bandwidth and integration density. As a result, extra constraints should be set in the mapping.
Enhancing the data locality can help avoid long-distance data transfer, which can assist in satisfying these constraints.


\subsubsection{Architecture Level}
Traditional large-scale deep learning acceleration platforms usually have clear control and communication boundaries, so the scaling out on them would suffer from the frequent host-to-host and host-to-device communication, and the significant gap between intra-device and inter-device bandwidth, leading to too coarse-grained task mapping and underutilization of hardware resources.
On the contrary, wafer-scale chips adopt a seamless architecture and basically eliminate the bandwidth gap between intra-die and inter-die, so they support uniform fine-grained mapping.
Owing to this, wafer-scale chips have much more potential in mapping flexibility and scalability to achieve better hardware utilization and overall performance, however, at the cost of much higher problem complexity of compilation.
Take the compilation in Figure \ref{fig:CUPOker1} as an example, the task can be divided and mapped to kernels with any shapes and executed in any regions of the wafer (if only the three constraints are satisfied).
If the target platform is a traditional accelerator cluster, the task division will be far more constrained by the boundary of devices.
To achieve such fine-grained search, traditional deterministic search methods such as dynamic programming \cite{bellman1966dynamic} may result in excessively long search times. In order to improve efficiency, heuristic search methods such as simulated annealing\cite{kirkpatrick1983optimization}, ant colony optimization\cite{dorigo2006ant} and genetic algorithms\cite{forrest1996genetic}, or even reinforcement learning\cite{thrun2000reinforcement}-based search may need to be developed.

\subsubsection{System Level}
The disruptive Wafer-scale Computing paradigm brings unprecedented challenges in building a computing system.
To address these challenges, researchers have to propose targeted hardware design strategies.
The compilers should also be aware of the hardware details and cooperate with them to achieve a satisfactory system optimization result.
For example, redundant process elements and data paths are designed to solve the yield problem, the compilers should know when and where to invoke the redundancies.
Another example, wafer-scale chips suffer from more serious thermal problem than traditional accelerators, the compilers can integrate hotspot-aware strategies in the original load balance procedure to complement the hardware cooling design.


\section{Integration \label{secInteg}}


As mentioned in the Section \ref{secBkg}, chiplet technology integrates multiple small dies into large computing systems by three main advanced packaging types: 
substrate-based, silicon-based, and redistribution layer (RDL)-based
packaging technologies.
In order to achieve the integration of wafer-scale chips, researchers have made optimizations on these original packaging technologies, or even take a more aggressive approach to directly produce an integrated wafer-scale chip.
In the rest of this section, we will introduce different integration types for wafer-scale chips, and give some discussions.

\subsection{Silicon-Based Integration \label{secIntegSi}}

The improvements of integration density and performance of large scale integration devices are mainly based on increasing the total number of input/output and power/ground terminals, which leads to shrinking design rule of wiring and bump pitch.
It is difficult to obtain highly reliable connections between chip and organic substrate with smaller bumps due to the mismatching of the coefficient of the thermal expansion\cite{ECTC00SI}.
To overcome these problems, silicon interposer-based packaging technology\cite{ECTC00SI} was proposed, where the interconnection between dies is implemented by an extra silicon layer between the substrate and die, and the connection between die and substrate is implemented by through-silicon vias (TSVs) and micro-bumps.
Since the micro-bumps and TSVs have smaller bump pitch and trace distance, silicon interposer provides higher IO density, lower transmission delay and power consumption than organic substrates~\cite{ChipletSurvey1}.


However, silicon interposers are finally connected to organic substrates, adding an extra level in the packaging hierarchy, which leads to a limited size and high overall packaging cost \cite{ECTC18SiIF2}.
To support wafer-scale integration, Bajwa et al. develop a package-less silicon-based integration platform named silicon interconnect fabric (Si-IF)\cite{ECTC17SiIF,ECTC18SiIF2,ECTC18SiIF}.
Si-IF supports heterogeneous reliable integration of chiplets with high interconnect density ($4 \times 10^6 cm^{-2}$), low interconnect resistivity ($0.8–0.9\Omega-um^2$), close inter-chiplet spacing ($\leq 100 um$), high adhesion strength ($150 MPa$), and with more uniform heat spreading\cite{ECTC18SiIF2,ECTC18SiIF}.
Si-IF does not require organic substrates, so the packaging cost is reduced.
Based on Si-IF platform, Pal et al. from UCLA and UIUC propose a wafer-level processor system architecture which can integrate at most 2048 pre-tested chiplets on a passive silicon-interconnect wafer with a total area of about 15,000 mm$^2$.



\subsection{Redistribution Layer (RDL)-Based Integration \label{secIntegInFO}}

Besides silicon-based integration, another one key technology for physical implementation of die interconnection in chiplets is redistribution layer (RDL) -based fan-out packaging \cite{ECTC06FO,IEMT06FO}.
This technology eliminates wirebonding or wafer bumping and leadframe or package substrate.
Alternatively, it requires an RDL to carry the corresponding metal wiring pattern.
Here the "fan-out" indicates that the IO ports of chips are rearranged on the loose area outside the die.
Since the IO ports are not confined to the range of die, the average circuit length can be reduced to improve the signal quality, and the die area can be reduced to improve the integration density without worrying the space for placing IO ports \cite{ChipletSurvey1}.

\begin{figure}
	\centering
	\includegraphics[width=.95\linewidth]{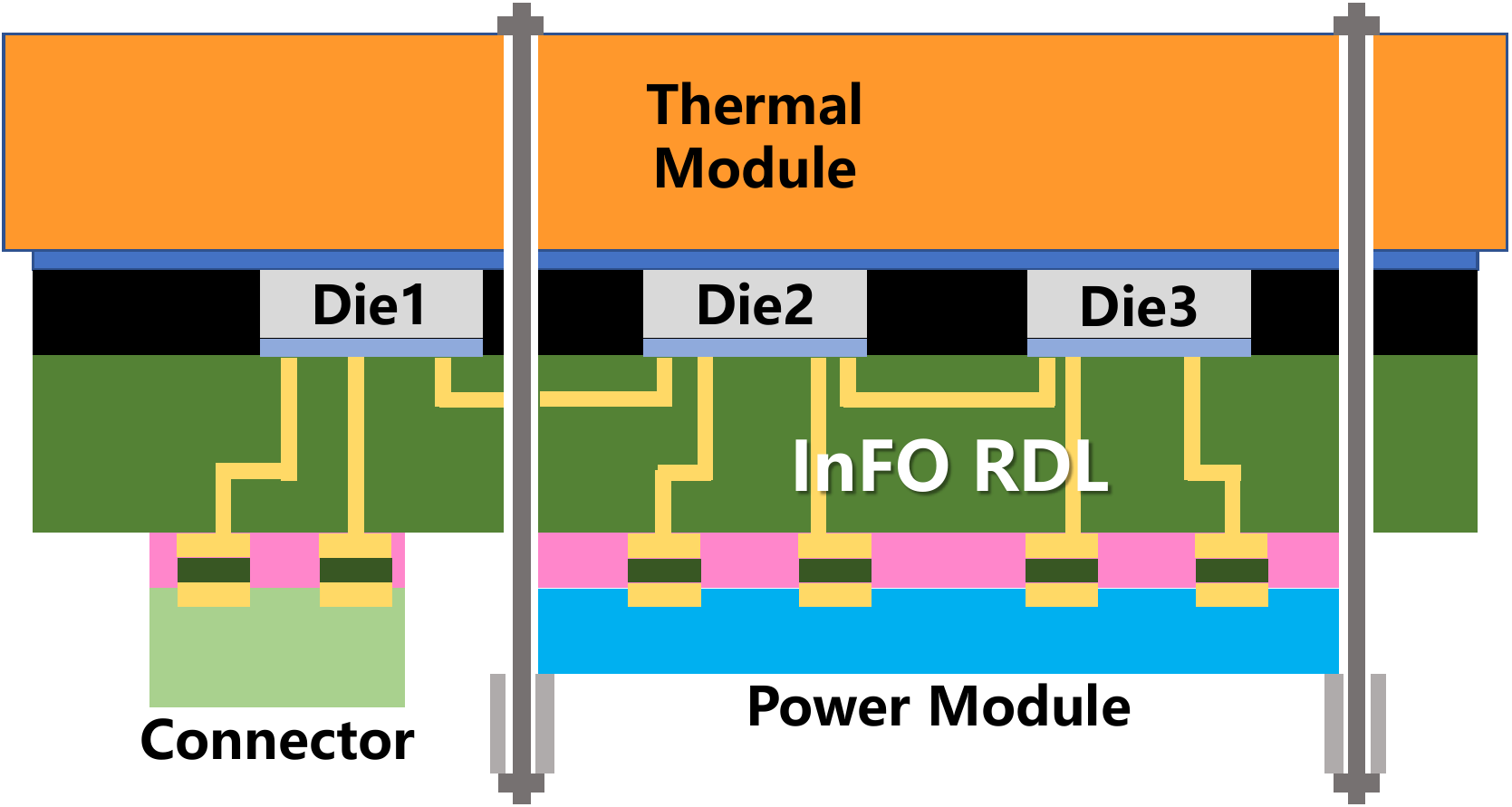}
	\caption{Schematic cross section of InFO\_SoW structure. Adapted from \cite{ECTC20InFOSoW}.}
	\label{figIntegInFOSoW}
\end{figure}

TSMC proposes the industry’s first wafer-scale system integration package with InFO technology, called InFO\_SoW \cite{ECTC20InFOSoW}.
As shown in figure \ref{figIntegInFOSoW}, an InFO\_SoW-based wafer-scale system integrates InFO, power and thermal modules.
Connectors and power modules are solder joined to the InFO wafer, which is then followed by the assembly of thermal module.
InFO\_SoW eliminates the use of PCB substrate by serving as the carrier itself. 
It has revealed good process uniformity across the super large package.
Lower surface roughness of InFO RDL saves about 15\% power of the interconnects with length of 30mm.
The 2-by-5 array dummy heater can dissipate 7000W while keeping the temperature below 90$^{\circ}C$.
Moreover, InFO\_SoW has relatively low risk on structural robustness when compared to qualified Flip-Chip package.
Tesla Dojo \cite{HCS22Dojo} is the first commercial product using InFO\_SoW technology, of which the area of a single chip (training tile) is over 16,125 mm$^2$.

\subsection{Field Stitching-Based Integration\label{secIntegFieldStitch}}

Unlike the above types which utilize advanced packaging technology\cite{Book21AdvPack} to integrate regular-sized dies together, Cerebras takes a more aggressive approach.
It uses the field stitching technology\cite{ICWSI92Lithographic,MICRO21CBStory} to connect the reticles, so that to directly produce an integrated wafer-scale chip, of which the area reaches as high as 46,255 mm$^2$.
This kind of wafer-scale integration provides high-capacity efficient SRAM distributed on the wafer, and ultra short inter-die links leading to uniform bandwidth across entire wafer.

In 1980, Trilogy Systems attempted to produce a single-chip 2.5 inches on a side\cite{MICRO21CBStory}.
Since the reticle limit constrained chip size to a maximum of 0.25 inch on a side at that time,
Trilogy Systems had to use transistor geometries larger than the minimum available, which sacrificed transistor density and impeded performance improvement\cite{MICRO21CBStory}.
To avoid this problem, Cerebras adopts the standard exposure process and field stitching (also called reticle stitching) technology \cite{ICWSI92Lithographic}.
First, a standard reticle of 525 mm$^2$ is repeated to traverse and expose the full wafer.
Then, an offset reticle of wiring between the 525-mm$^2$ reticles is used to “stitch” the standard reticles together\cite{MICRO21CBStory}.
The process of field stitching was published decades before\cite{ICWSI92Lithographic},
but this is the first attempt at applying it to a commercial wafer-scale integration product.

\subsection{Discussion \label{secIntegDisc}}



Since advantaged packaging-based integrations allow using individual pre-tested chiplets (i.e., known-good-dies, KGDs) for assembling into a wafer-scale chip, they can mitigate the die yield problem to a large extent.
On the contrary, Cerebras directly produces an integrated wafer-scale chip, so the die yield challenge is more significant.
To address it, Cerebras implements a homogeneous array of small cores, and reserves about 1\% redundant cores in order to “repair” defective ones \cite{HCS19Cerebras,MICRO21CBStory}.

\begin{figure}
	\centering
	\includegraphics[width=.95\linewidth]{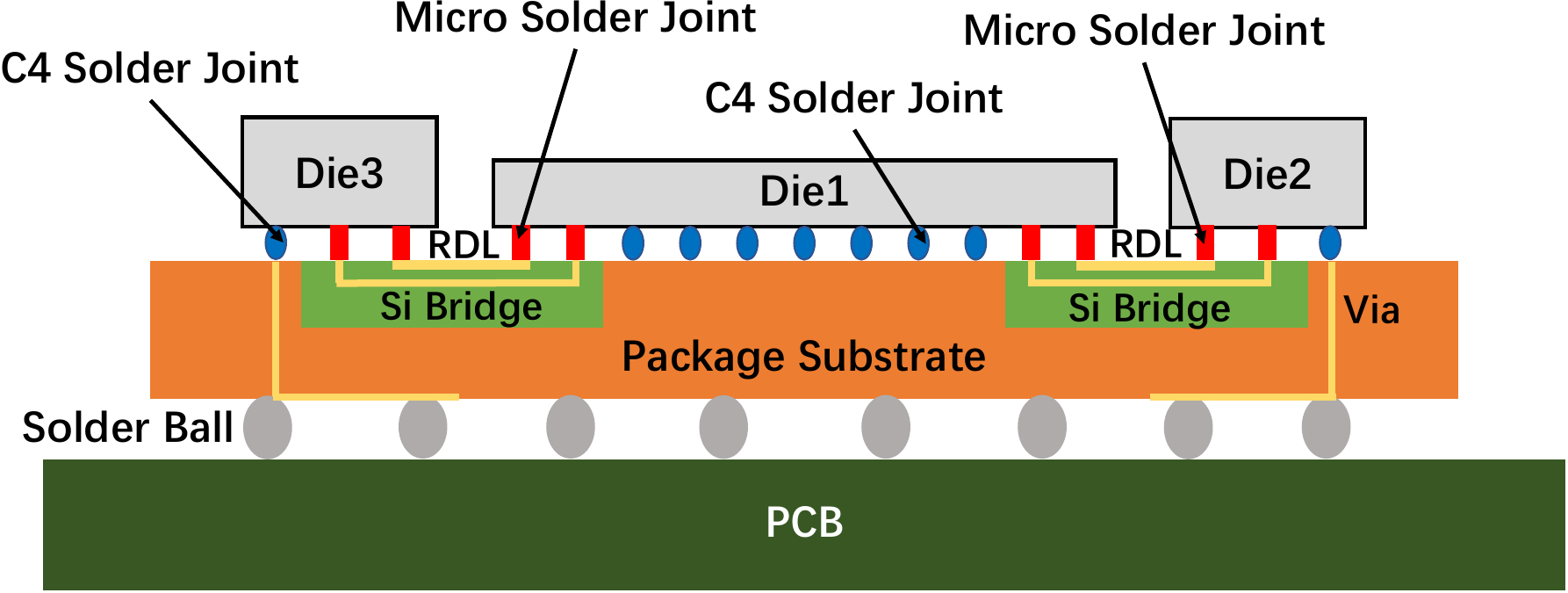}
	\caption{Chiplet heterogeneous integration on organic substrate with silicon bridge. Adapted from \cite{Book21AdvPack}.}
	\label{figIntegSiliconBridge}
\end{figure}

Compared to the organic substrate, the implementation of silicon interposer brings higher cost in materials and process.
The cost of RDL-based fan-out integration falls in between the organic substrate and silicon interposer, but the wiring resources of fan-out packaging are limited by the RDL wiring level.
Another path to reduce the cost of silicon interposer is silicon bridge technology \cite{ECTC16EMIB,ECTC18SiBridge}.
As shown in figure \ref{figIntegSiliconBridge}, it integrates small thin layers on the substrate (instead of a complete silicon interposer) for inter-die interconnection.
Since silicon bridge technology makes a good balance between performance and cost,
multiple industries have carried out study on it and proposed the corresponding chiplet products,
such as Intel \cite{ECTC16EMIB}, Apple \cite{Micro22AppleM1} and TSMC \cite{HCS21TSMCPack}.
However, the sizes of these products are far from reaching the scale of wafer.

Field stitching (or reticle stitching) is a fundamental technology proposed in the last century \cite{ICWSI92Lithographic}.
One main disadvantage of reticle stitching is that the mask cannot be change during a complete exposure on the full wafer, otherwise the accuracy would seriously decrease.
That means a great challenge for heterogeneous integration by reticle stitching.
Contact printing \cite{contactPrint1997,contactPrint2001,contactPrint2007,contactPrint2022} supports more free designs, and may become an inspiring research direction for wafer-scale integration if its defects problem can be solved.



\section{System \label{secSystem}}

In this section, we will introduce three key design issues of building a wafer-scale system: 
power delivery, clock delivery and heat dissipation.

\subsection{Power Delivery \label{secSystemPower}}
For wafer-scale chips, stable power transmission is inseparable from the high computing power. 
How to stably transmit the power supply voltage to each die in the wafer chip is a question worth considering. 
Currently, there are two main styles of power delivery for wafer-scale chips:
1) Edge (Lateral) style: deliverying power through the edge of the wafer, and
2) Vertical style: using the third dimension to power the entire wafer.



\subsubsection{Edge style}
UCLA\&UIUC's work \cite{2048chiplet} adopts edge power delivery, because its integration of through-wafer-vias technology a Si-IF wafer \cite{8811403} is still under development and not mature yet.
According to different actual needs, two strategies are provided \cite{2048chiplet}: 
1) Delivering high voltage (say 12V) power at the edge and using down conversion near the die \cite{pal2019architecting}. 
This method will cause a huge area overhead (25\%-30\%) due to the need to place huge components such as inductors and capacitors on the wafer. In addition, these additional components will destroy the chain of the die array circuit and increase the distance between dies, making the overall design more complex. 
2) Delivering higher voltage supplies (say 2.5V) at the edge and using low dropout (LDO) based regulation in the die, which also means some power efficiency loss in the power delivery plane due to resistive power loss and poor LDO efficiency.
In addition, since the power is delivered from edge to center, one fault in the chain may cause large-scale failure.

\subsubsection{Vertical style}
Different from UCLA\&UIUC's work, a vertical power supply program is provided by Tesla in \cite{DojoPatent1}. 
A two-layer voltage regulating module (VRM) is designed.
The first layer is configured to output a regulated voltage based on a stepped down voltage,
and the second layer includes a plurality of active components configured to provide the stepped down voltage to the first layer.
Vertical structure brings a better thermal budget. 
Enhancing the hardware architecture stability, BGA is used to connect the layers and its clearance can be designed as 1.1 mm +/- about 3\% depending on the implementation. 
Delivering about 600 watts of power and a range of 0.8-1.1 volts of voltage relying on the embodiment, this module offers sufficient power and voltage for high-performance computation.

Cerebras provide another vertical architecture \cite{CerebrasPatent1}. Between two surfaces, an integrated circuit has a first set of conductive pads arranged along the second surface of the circuit opposing the first surface, while a power converter is electrically connected to each of the first set of conductive pads, playing the role of adjusting the voltage in a suitable range. Different from BGA, the conductive pads perform better in the variant working situation, where it can choose suitable Young’s Modulus (1N/m-1MN/m) alternatively after preload compression. Addressing the problem of hardware mismatch, this flexible property benefits stability in different work situations and different hardware parameters. The power components can accept any suitable AC and/or DC input voltage and it can output any suitable voltage, such as 0.9 VDC or 0.8 VDC.

In conclusion, when compared to edge power supply, the vertical architecture showcases several advantages in terms of output power, stability, and flexibility. Nevertheless, it does entail higher costs and necessitates a higher level of technical maturity.


\subsection{Clock Delivery \label{secSystemClock}}
For a wafer-scale chip, how to reliably distribute clock across such a large area is a challenge. 
Similar to power delivery, there are also edge and vertical styles of clock distribution for wafer-scale chips.

\subsubsection{Edge style}
UCLA\&UIUC's work \cite{2048chiplet,pal2021scale} has presented an edge-style solution, which builds a clock distribution network, generates a fast clock (up to 350 MHz) in one edge die from a slower system clock provided by an off-die crystal oscillator source, and then forwarded throughout the die array by forwarding circuitry built inside every die.
To avoid pull-up/pull-down imbalances in I/O drivers between various components and chips, the design also propagates an inverted version of the clock, ensuring that the distortion alternates between half the clock cycle.
A faulty die can disrupt the clock propagation mechanism. Both the generation and propagation of the clock require a fault tolerance design. In UCLA\&UIUC's work, each die receives the rebroadcast clock of the adjacent die, and each non-edge block receives the propagation clock from all four directions, which reduces the probability of die failure and improves the resiliency of clock propagation.

\subsubsection{Vertical style}

Edge style clock distribution is a reasonable solution for TSV-less structures like UCLA\&UIUC's Si-IF-based chip. However, long-distance clock transmission can lead to extra clock jitter, energy consumption, heat generation and parasitic inductance.
To avoid these problems, Tesla proposed a vertical clock scheme
\cite{TeslaPatentClock}
. They transmit the clock and power to the die from a vertical direction, so that long-distance transmission can be avoided.
However, due to the limited area of the PCB board shared by the clock and power, the VRM is closer to the clock circuit, and the clock may be contaminated by power noise.
In order to address this issue, Tesla suggests a series of techniques, including differential signaling clock, filter circuit, isolated ground wire, and so on.
Overall, the vertical clock scheme presents higher technical hurdles and increased costs; however, it holds significant potential for future development.

\subsection{Heat Dissipation}
Heat dissipation is extremely important for wafer-scale chip, which directly affects the overall performance of the chip. Although the thermal design power (TDP) of a single system on wafer (SoW) could reach an astonishing order of 10$^4$W, the huge area of the SoW keeps its heat power density at a level comparable to the most advanced GPUs and TPUs. 
According to our investigation and projections, the heat power density of SoW is between 20$\sim$100 cm$^2$. 
However, the large area and the tremendous total power of SoW make it impossible to directly apply traditional chip cooling methods. Therefore, customized cooling solutions make essential, and emerging cooling technologies such as microfluidics \cite{wang20183d}, may also become one of the options for future SoW cooling solutions.

\subsubsection{Cerebras WSE}
the WSE series of Cerebras adopts an air-cooling system with an internal coolant loop in \cite{CS1Cooling}.
The coolant is water/propylene glycol mixture, which is transported in the loop at low pressure and high flow rates while remaining single phase, and takes heat away from the surface of the WSE through connectors. Meanwhile, the heat transferred by the coolant is dissipated through a set of fans and a heat exchanger. Although this method implements cooling $>$ 15kW on a single WSE with an area of 46,625 mm$^2$, the huge heat dissipation system occupies more than 90\% of the volume of the chassis, which reduces the possibility of WSE chips forming a high-density computing system in units of wafers.

\subsubsection{UCLA\&UIUC's work}
UCLA\&UIUC’s wafer-scale system is cooled by forced air convection using one or two square radiators \cite{pal2019architecting}. Two heat sinks cover the wafer chips, one directly on top of the die and the other on the backside of the wafer. This structure not only provides mechanical support for the wafer, but also helps improve heat dissipation efficiency. In the extreme condition of not considering factors such as power supply, the cooling device of dual heat sink can provide the system with a TDP of 9300W in 50,000 $mm^2$ area, while keep the junction temperature below 120$^{\circ}C$. However, in the face of higher heat dissipation requirements, this method based on traditional air cooling will encounter a bottleneck.

\subsubsection{Tesla Dojo}
In \cite{DojoPatent1}, the thermal system of Tesla Dojo includes a thermal dissipation and a cooling system, which are distributed on both sides of the SoW and connected together by conductive frames distributed around the SoW to provide heat dissipation, as well as structural support and EMI shielding. The thermal dissipation connected to the SoW directly or through a thermal interface material with low heat transfer resistance, dissipate heat from the SoW and may consists of a metal platform and/or heat sink. The cooling system is installed above the VRMs that supply power to the SoW, providing active cooling for the VRM node and the SoW. It can take the form of customized liquid cooling, e.g., metal with flow paths, or include brazed fin arrays to increase the cooling efficiency. 
For the specific implementation of the cooling system, Shu-Rong Chun, et al. proposed an expandable liquid-cooling platform in \cite{ECTC20InFOSoW}, which consists of multiple sub-unit cold plates, each responsible for cooling a row of dies of SoW. The coolant is deionized water and optimally distributed throughout each sub-unit cold plate. The liquid-cooling platform enables a dummy heater array with a heat density of 120 cm$^2$ to be kept at a temperature of about 65$^{\circ}C$.

\subsubsection{Microfluidics}
Several current cooling technologies with practical applications, such as heat pipes or heat sinks as the mainstream cooling technology for server companies, single-phase liquid cooling technology applied to high-power servers, and two-phase immersion cooling used in supercomputers, can meet the cooling needs of the SoW level. However, the massive volume overhead of these technologies is not what SoW’s designers expect. In this context, we turn our attention to the emerging microfluidic, also called microchannels, which has the advantages of small device size, high heat transfer efficiency, low thermal resistance and fast heat transfer, and is recognized as a promising heat dissipation method consistent with SoW.
More recently, a mainstream microfluidic technique is achieved by etching microchannels or microcolumns in the silicon intermediate layer, and then combining different silicon intermediate layers through silicon-silicon bonding, silicon -metal bonding or other processes, and using the forced flow of liquid coolant or two-phase coolant to carry away the heat of the chip\cite{wang20183d,zheng2015silicon,zheng2013design}. Due to its small enough size and high efficiency from the active cooling, microfluidics has been considered as promising for 3D ICs with high power density \cite{wang20183d}. The experimental result in \cite{zheng2015silicon} shows that the temperature of 3D chip stack with heating density of 97 $cm^2$ can be controlled below 55.9 $^{\circ}C$ under the condition of total flow of 50mL/min. In addition, microchannel can provide heat dissipation of close to 1000 $cm^2$ at sufficiently high flow rates and pressure drops \cite{tuckerman1981high}. Meanwhile, it is also important to note that microfluidics are suitable for almost any IC with a silicon intermediate layer structure, including the SoW. Compared with other heat dissipation methods, the sufficiently small geometry, outstanding heat dissipation performance, and wide range of application scenarios of microfluidic technology make it highly likely to become the next generation of disruptive IC cooling technology. However, the complex manufacturing process and high price have always been the most difficult obstacles to overcome on the road to its commercialization.



\section{fault tolerance\label{secFault}}

In previous sections, we have interspersed fault tolerance designs from different perspectives.
As fault tolerance is crucial for Wafer-scale Computing, we will now provide a summary of these designs in this section.

\subsection{The Necessity of Fault Tolerance Designs}

The yield of chip production refers to the proportion of chips that meet the specifications and quality requirements out of the total output during the production process \cite{Yield1986,Yield2006,Yield2008,Yield2010}.
It is primarily influenced by the following factors: die area, process, defect density, and other practical factors.
The yield is an important indicator in chip production, as it directly affects the cost and performance of chips.
Furthermore, it has been a key factor constraining the development of the chip industry.

Wafer-scale chip production typically faces significant yield challenges.
Taking Cerebras' WSE-2 \cite{HCS22Cerebras} as an example, each die is characterized as 510 mm$^2$ area and TSMC N7 7nm process with a defect density of 0.09 defects per square centimetre \cite{defectDensity}.
According to the yield models such as the Murphy's yield model \cite{MurphYieldModel1,MurphYieldModel2,yieldCalculator} or the Chiplet Actuary model \cite{feng2022chiplet},
the estimated die yield is only around 60\%.
This implies that approximately 40\% of the dies have at least one defect and cannot function properly without specialized optimization.

Advantaged packaging allows using individual pre-tested chiplets (KGDs) for assembling into a wafer-scale chip, so that mitigates the die yield problem to a large extent.
However, defects may also occur in the interconnections, and the bonding defects lead to waste of KGDs \cite{feng2022chiplet}.
Research conducted by UCLA\&UIUC's \cite{2048chiplet} reveals that though the bonding yield per I/O can exceed 99.99\%, the overall bonding yield for over 2000 I/Os per chiplet is only 81.46\%.
Moreover, even with an improved bonding yield of 99.998\% per chiplet, the overall system of 2048 chiplets might still have a few faulty chiplets.
Therefore, fault-tolerance design is essential for Wafer-scale Computing systems.

\subsection{Summary of Fault Tolerance Designs} 

\subsubsection{Computation Core}
To deal with the yield problem of computation core, Cerebras adopts a small core design and reserves redundant cores for replacing the faulty ones \cite{HCS19Cerebras,lie2022processor}. On the other hand, Tesla integrates KGDs into a wafer-scale chip \cite{HCS22Dojo}, so the yield problem of dies is under control, at the cost of a larger bandwidth gap across the die boundary than Cerebras' stitched wafer (but still much smaller than traditional accelerator clusters). These are discussed in Section \ref{secArchDisc}.

\subsubsection{Interconnection}
Besides the computation cores, faults can also happen in the interconnections.
To improve the I/O bonding yield, UCLA\&UIUC's work suggests landing two copper pillars on each I/O pad  \cite{2048chiplet}.
Cerebras reserves redundant fabric links for reconnecting fabric and restoring logical 2D mesh \cite{HCS19Cerebras,lie2022processor}.
Naturally, appropriate protocols should be designed to activate the backup links.
To reduce the impact of faulty computation cores on interconnections,
the NoC in a wafer-scale chip is usually designed to be highly decoupled from computation cores
 \cite{HCS19Cerebras,lie2022processor}. 
Moreover, the topology of NoC also influences fault tolerance.
2D Mesh and Torus are common-used in existing wafer-scale chips, because they can provide good support for fault tolerance designs (e.g., X-Y/Y-X dimension-ordered routing \cite{2048chiplet}) with small physical implementation difficulty. These are discussed in Section \ref{secArchNoC}, \ref{secArchDisc} and \ref{secInterconnectDisc}.

\subsubsection{Power and Clock}

For power and clock supply, edge-style schemes require a chain or network for delivering the power/clock from edge to center.
Consequently, a single fault can lead to widespread failure. 
Therefore, redundant paths must be designed to reduce the affected area.
In contrast, vertical-style schemes inherently offer fault tolerance.
However, these schemes also come with higher technical hurdles and costs now.
These are discussed in Section \ref{secSystemPower} and \ref{secSystemClock}.

\subsubsection{Compilation}

If the local controller and protocol are properly designed, the local fault tolerance operations (e.g., activating the backup logic/link in a same core when the main logic/link is broken) can be automatically done without the need to notify the upper-layer applications.
However, to achieve better fault tolerance effect, the compiler should perform fault tolerance optimizations from a global perspective.
One possible solution is to run a test program when the hardware system is started, characterize the hardware system to let the compiler know which parts are broken and where the backup redundancies can be used.
Taking this information into consideration, the compiler maps the workload properly, and is prepared for adjusting the mapping if new hardware failures appear at runtime. These are discussed in Section \ref{secToolDisc}.


\section{application\label{secApp}}

In the previous sections, we use AI computation task, specifically neural-network-based deep learning acceleration, as the typical application scenario. However, Wafer-scale Computing is not limited to deep learning, and can be applied to various other tasks. 
In this section, we will introduce the important scientific computing applications~\cite{Powering} 
of Wafer-scale Computing system outside of AI computing domain. These applications leverage the architecture's high on-chip bandwidth and fast memory access, which are challenging to achieve with traditional accelerator clusters.

\subsection{Linear Algebra}
Linear algebra~\cite{LA} is a fundamental application in the field of High-Performance Computing (HPC), as many engineering and scientific problems require transformation into linear algebra problems to be solved. Even neural network computation can be seen as basic matrix calculations. The application of linear algebra requires large-scale parallel computing, both dense and sparse. Wafer-scale Computing systems boast high computing power density, and the on-chip distributed SRAM can efficiently handle data access requests, resulting in significantly improved parallel performance.

\subsection{Stencil Computation} 
Stencil computation, or Finite Element Methods~\cite{DBLP03660}, is widely used in Partial Differential Equation (PDE) applications. PDEs are mathematical equations describing various phenomena in nature, typically involving derivatives of spatial and temporal variables. Numerically solving PDEs requires discretization of continuous physical domains (e.g., spatial regions) into grids or arrays, approximating the differential operator of the PDE.
In these computational applications, the arithmetic intensity of single elements is low, which makes traditional hierarchical memory systems unsuitable. On the contrary, in Wafer-scale Computing system, each processor equipped with dozens of kB SRAM can be assigned an appropriate stencil kernel. The cores can efficiently exchange data with neighbors through a high-bandwidth fabric.

\subsection{N-body Problems}
The N-body problem~\cite{N-body, N-bodyppt} involves calculating the interactions between every pair of particles, where each particle is influenced by gravity or interaction forces from all other particles. As the number of particles (N) increases, the calculation amount grows sharply, requiring N$^2$ interaction calculations. N-body computations involve complex communication patterns between particles. Thanks to their flexible data routing and single-cycle memory access capabilities of on-chip SRAM, Wafer-scale Computing systems perform excellently in these N-body applications.

\subsection{Spectral Methods}
Spectral Methods~\cite{Spectral-Methods} are numerical computing applications commonly used for solving PDEs and other mathematical problems. This method expresses the function as a linear combination of orthogonal basis functions. By computing on these basis functions, the original PDEs can be transformed into algebraic equations, and numerical solutions can be obtained by solving the resulting algebraic equations.
Due to the global nature of this application, it exhibits typical all-to-alls and reductions communication behavior, posing challenges for traditional hardware architectures. Wafer-scale Computing systems can address these specific communication behaviors by their high-bandwidth and low-latency fabric between PEs.


\section{Conclusion\label{secCon}}

Wafer-scale Computing is an emerging trend to solve the gap between the computing powers required by large artificial intelligence models and provided by the chips.
Existing publications mainly introduce individual products and techniques, so there is an urgent need to create a comprehensive survey on the current state of knowledge of Wafer-scale Computing technologies.
We propose this paper to compare the different designs, summary their similarities and differences, extract the essential points of Wafer-scale Computing, discuss the achievements and shortcomings of existing research and give advice on the possible future research directions.
\section*{acknowledgement\label{secAck}}
This work was supported in part by NSFC Grant 62125403; in part by the Science and Technology Innovation 2030 - New Generation of AI Project under Grant 2022ZD0115200; in part by Beijing Municipal Science and Technology Project Grant Z221100007722023; in part by the National Key Research and Development Program under Grant 2021ZD0114400; in part by Beijing National Research Center for Information Science and Technology; in part by the Beijing Advanced Innovation Center for Integrated Circuits; and in part by Tsinghua University-China Mobile Communications Group Co.,Ltd. Joint Institute.



\bibliographystyle{IEEEtran}
\bibliography{ref/IEEEabrv,ref/refs}

\begin{IEEEbiography}[{\includegraphics[width=1in,height=1.25in,clip,keepaspectratio]{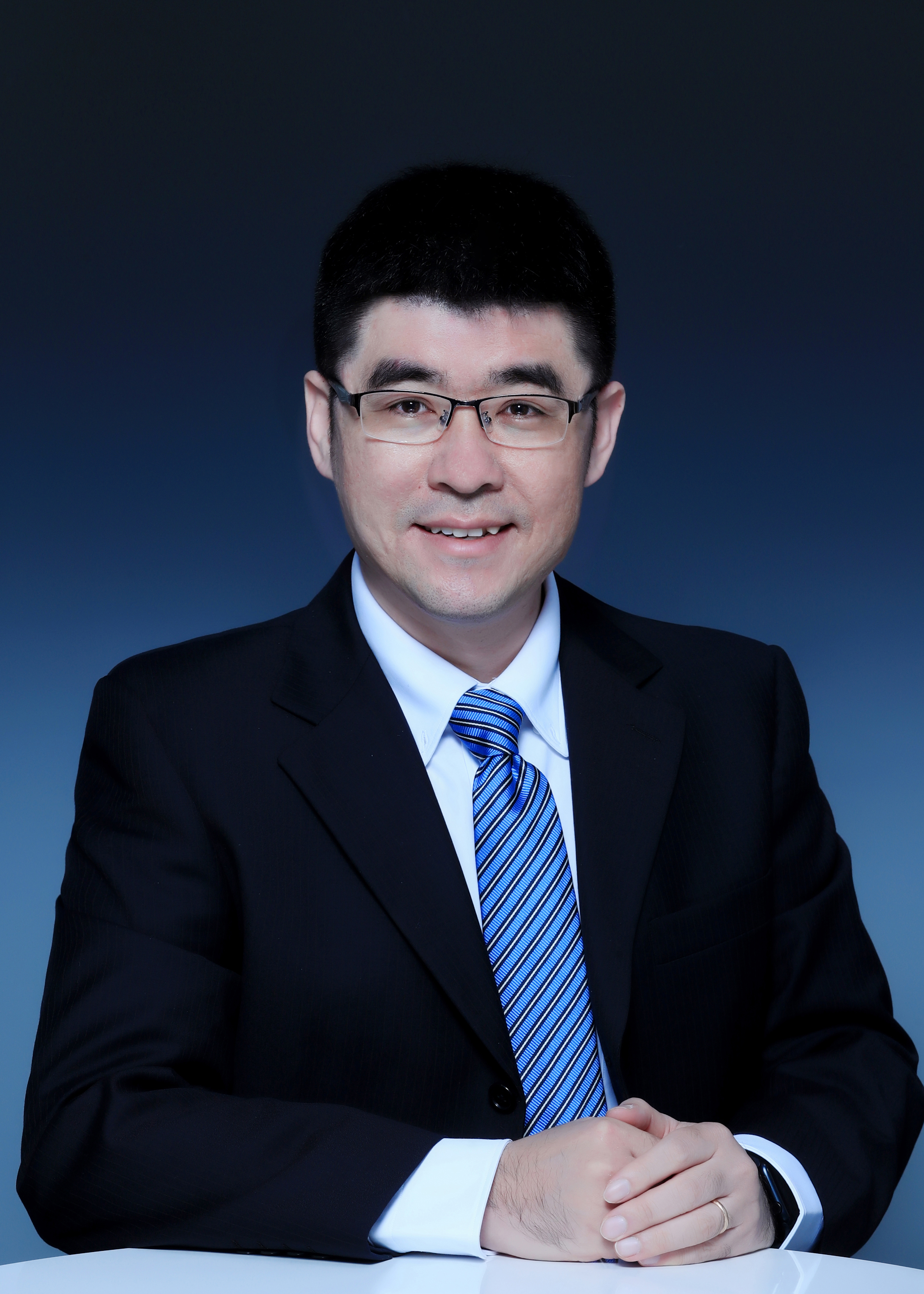}}]{Dr. Yang Hu} is currently an Associate Professor in the School of Integrated Circuits at Tsinghua University. He receives his BS degree from Tianjin University in 2007, and his MS degree from Tsinghua University in 2011, and his Ph.D. degree from the University of Florida in 2017. He was a tenure-track assistant professor in the ECE department at University of Texas at Dallas from 2017 to 2021. Dr. Hu now works on high-performance AI chip architecture and compilation tools.

Dr. Hu is an NSF CAREER Awardee. He has published more than 50 papers in computer architecture conferences and journals such as ISCA, HPCA, ASPLOS, MICRO, SC, DAC, ICS, RTAS, ICPP, ICCD, IEEE-TCAD, IEEE-TC, and IEEE-TPDS. His research work has won Best Paper Nomination of HPCA in 2017 and 2018. He received the Best of IEEE Computer Architecture Letters Award in 2015. He is an Associate Editor of Elsevier Chip Journal. He served as TPC track chair of DAC, and TPC member of HPCA, DAC, IWQoS, ISPASS, ICPP, ICDCS, IPDPS, and etc.. He served as NSF panelists and external reviewer of Hongkong Research Grant Council. He also served as session chair of HPCA 2022 and registration chair of ICS 2018.
\end{IEEEbiography}
\vspace{-20 pt}

\begin{IEEEbiography}[{\includegraphics[width=1in,height=1.25in,clip,keepaspectratio]{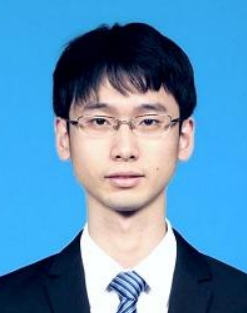}}]{Xinhan Lin}received the B.S. degree in computer science and technology from Sichuan University, Chengdu, China, in 2010, the MS degree in integrated circuits engineering from Tsinghua University, Beijing, China, in 2016, and Ph.D. degree in electronics science and technology from Tsinghua University, Beijing, China, in 2022.

Currently he is working for the Shanghai Artificial Intelligence Laboratory, Shanghai, China. His research interests include deep learning, reconfigurable computing and neural network acceleration.
\end{IEEEbiography}
\vspace{-20 pt}

\begin{IEEEbiography}[{\includegraphics[width=1in,height=1.25in,clip,keepaspectratio]{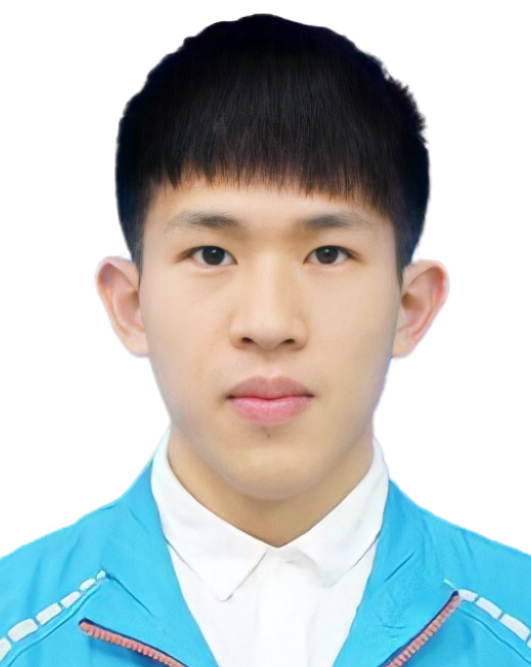}}]{Huizheng Wang} received the B.S. and M.S. degrees in communication engineering from Southeast University (SEU), Nanjing, China, in 2019 and 2022, respectively. He is currently working toward the Ph.D. degree at the School of Integrated Circuits, Tsinghua University, Beijing, China. His research interests include efficient algorithms and VLSI architectures for massive MIMO detection, polar decoder, stochastic computing, deep learning and neural network acceleration. 
\end{IEEEbiography}
\vspace{-20 pt}

\begin{IEEEbiography}[{\includegraphics[width=1in,height=1.25in,clip,keepaspectratio]{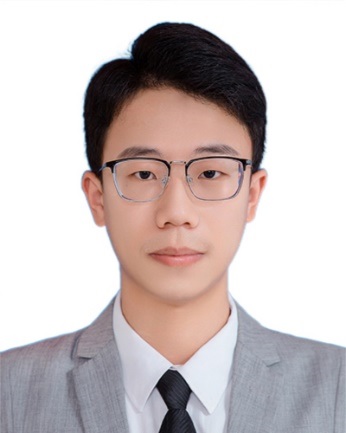}}]{Zhen He} received the B.S. degree from Chongqing University, Chongqing, China, in 2022.  He is currently pursuing the Ph.D. degree with the School of Integrated Circuits, Tsinghua University, Beijing, China. His research interests include deep learning, in-memory computing, and computer architecture.
\end{IEEEbiography}
\vspace{-20 pt}

\begin{IEEEbiography}[{\includegraphics[width=1in,height=1.25in,clip,keepaspectratio]{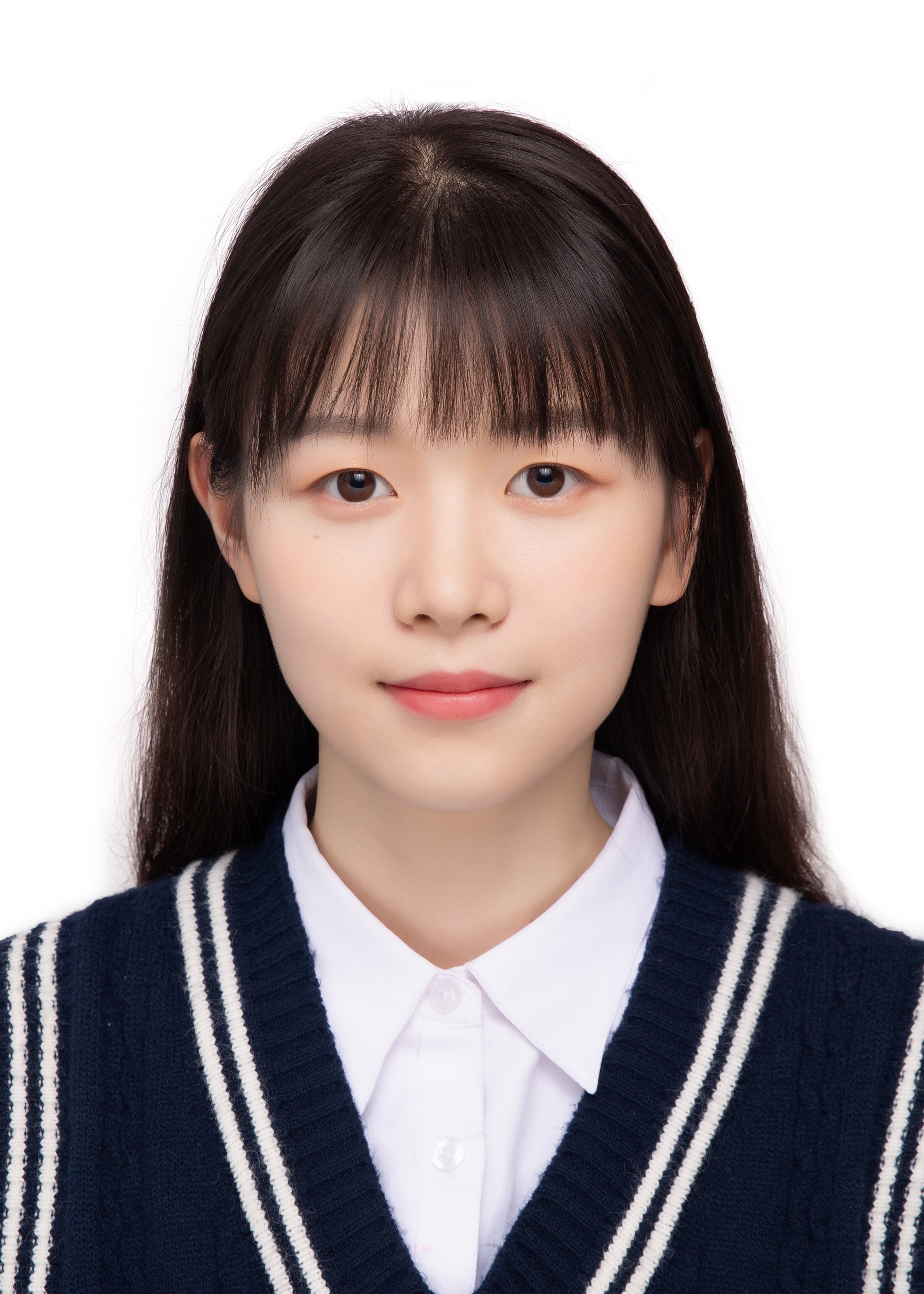}}]{Xingmao Yu} received the B.S. degree in computer science and technology from National University of Defense Technology, Changsha, China, in 2023, and she is working toward the M.S. degree in integrated circuit engineering from Tsinghua University, Beijing, China.
    
She's research interests include architecture design of chiplet-based systems, interconnect fabrics and system modeling.
\end{IEEEbiography}
\vspace{-20 pt}

\begin{IEEEbiography}[{\includegraphics[width=1in,height=1.25in,clip,keepaspectratio]{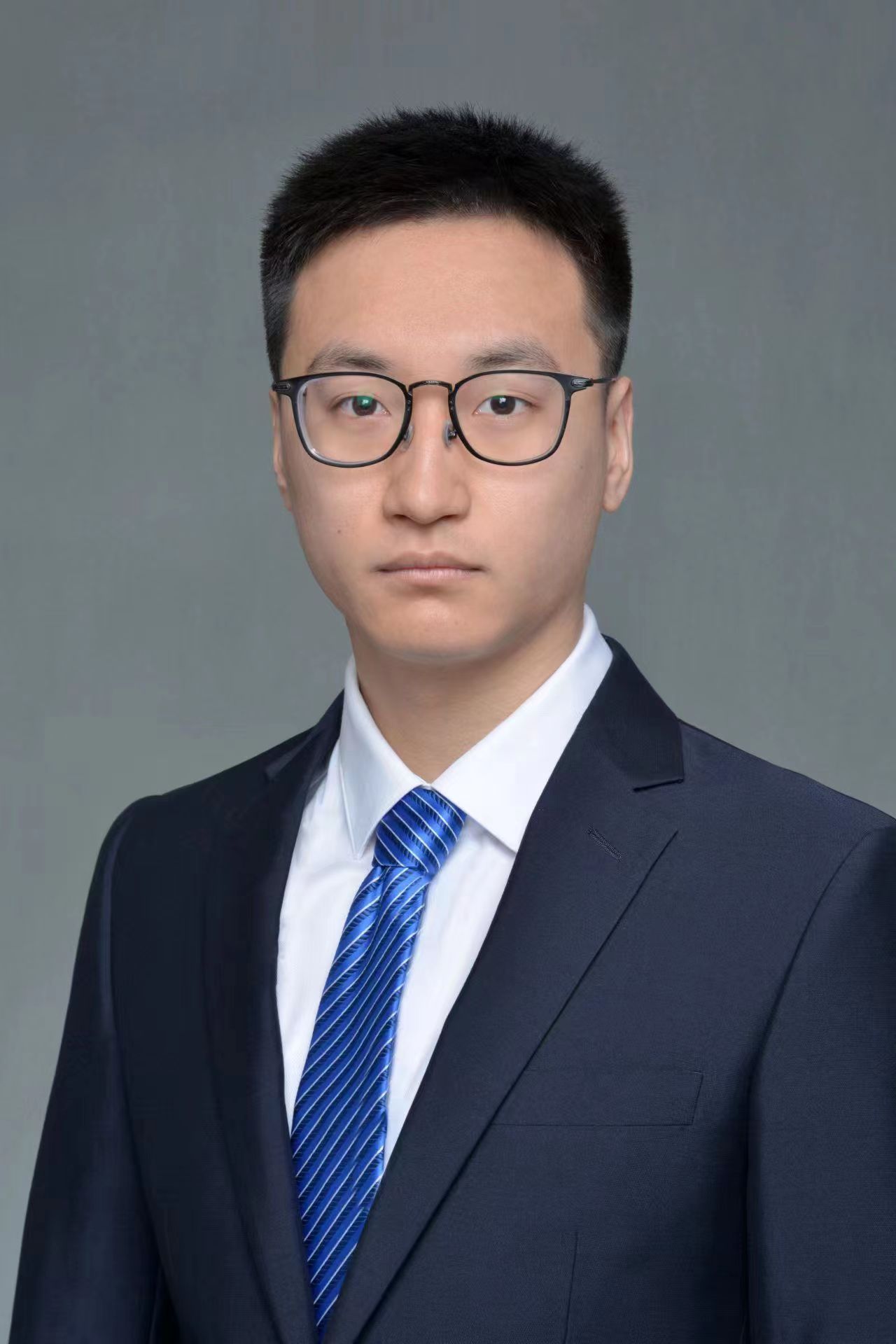}}]{JiaHao Zhang} receiced the B.S. degree in microelectronics science and engineering from the School of Intergrated Circuits, Tsinghua University, Bejing, China, in 2023.
        
He is currently working with Professor Shouyi Yin in the School of Intergrated Circuits, Tsinghua University. His research interests include computer architecture, AI acceleration and processors, and large-scaling chip design. 
\end{IEEEbiography}
\vspace{-20 pt}

\begin{IEEEbiography}[{\includegraphics[width=1in,height=1.25in,clip,keepaspectratio]{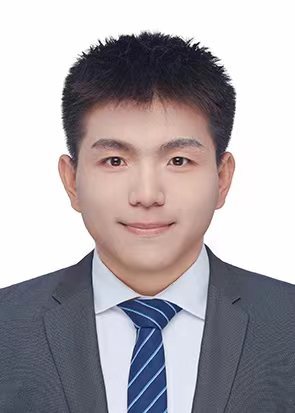}}]{Qize Yang} received the B.S. degree in micro electronic science and technology from Tsinghua University, Beijing, China, in 2022.

He is currently working toward the Ph.D. degree with the School of Integrated Circuits, Tsinghua unversity, Beijing, China. His research interests include distributed learning, wafer scale computing and communication, and neural network acceleration. 
\end{IEEEbiography}
\vspace{-20 pt}

\begin{IEEEbiography}[{\includegraphics[width=1in,height=1.25in,clip,keepaspectratio]{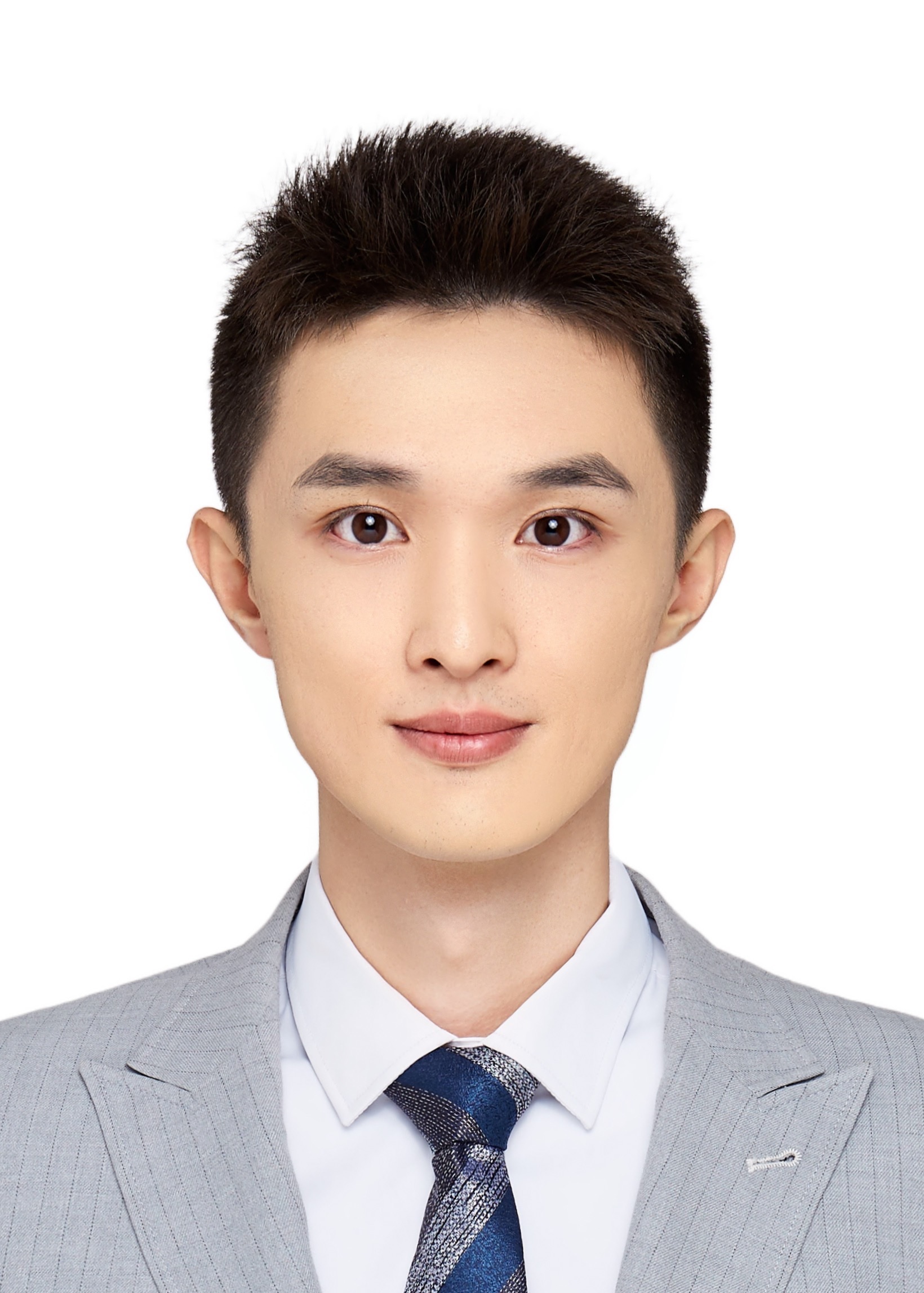}}]{Zheng Xu} received the B.S. degree from the School of Information and Electronics, Beijing Institute of Technology, Beijing, China, in 2022. 

He is currently a Ph.D. candidate with the School of Integrated Circuits, Tsinghua University, Beijing, China. His research interests include neural network acceleration, AI compiler, and wafer-scale chip.
\end{IEEEbiography}
\vspace{-20 pt}

\begin{IEEEbiography}[{\includegraphics[width=1in,height=1.25in,clip,keepaspectratio]{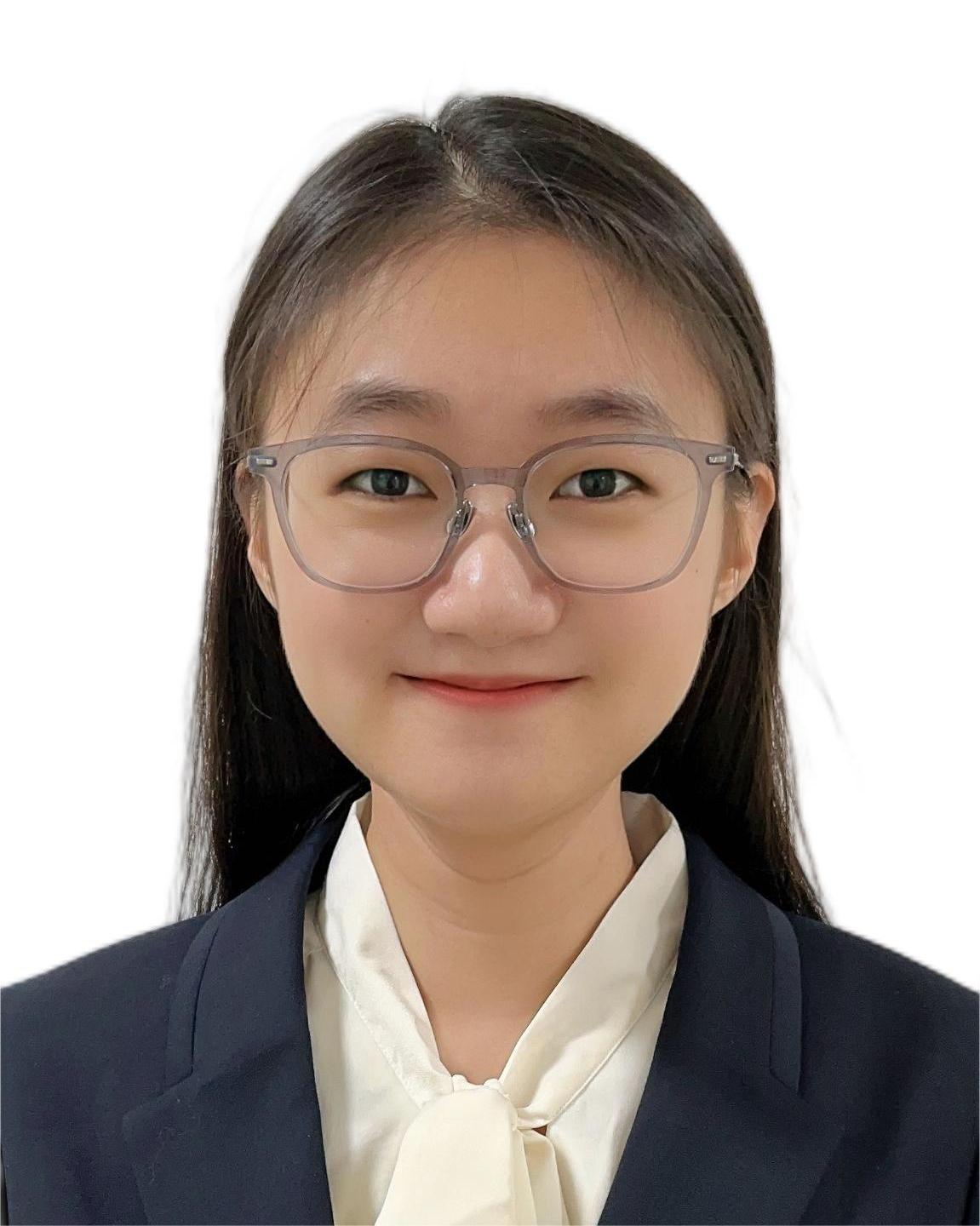}}]{Sihan Guan} received the B.S. degree in Electronics and Information Engineering from Beihang University, Beijing, China, in 2019. She is currently working toward the M.S. degree in Integrated Circuit Engineering with the School of Integrated Circuits, Tsinghua University, Beijing, China. Her research interests include neural network acceleration, hardware simulation and design space exploration.
\end{IEEEbiography}
\vspace{-20 pt}

\begin{IEEEbiography}[{\includegraphics[width=1in,height=1.25in,clip,keepaspectratio]{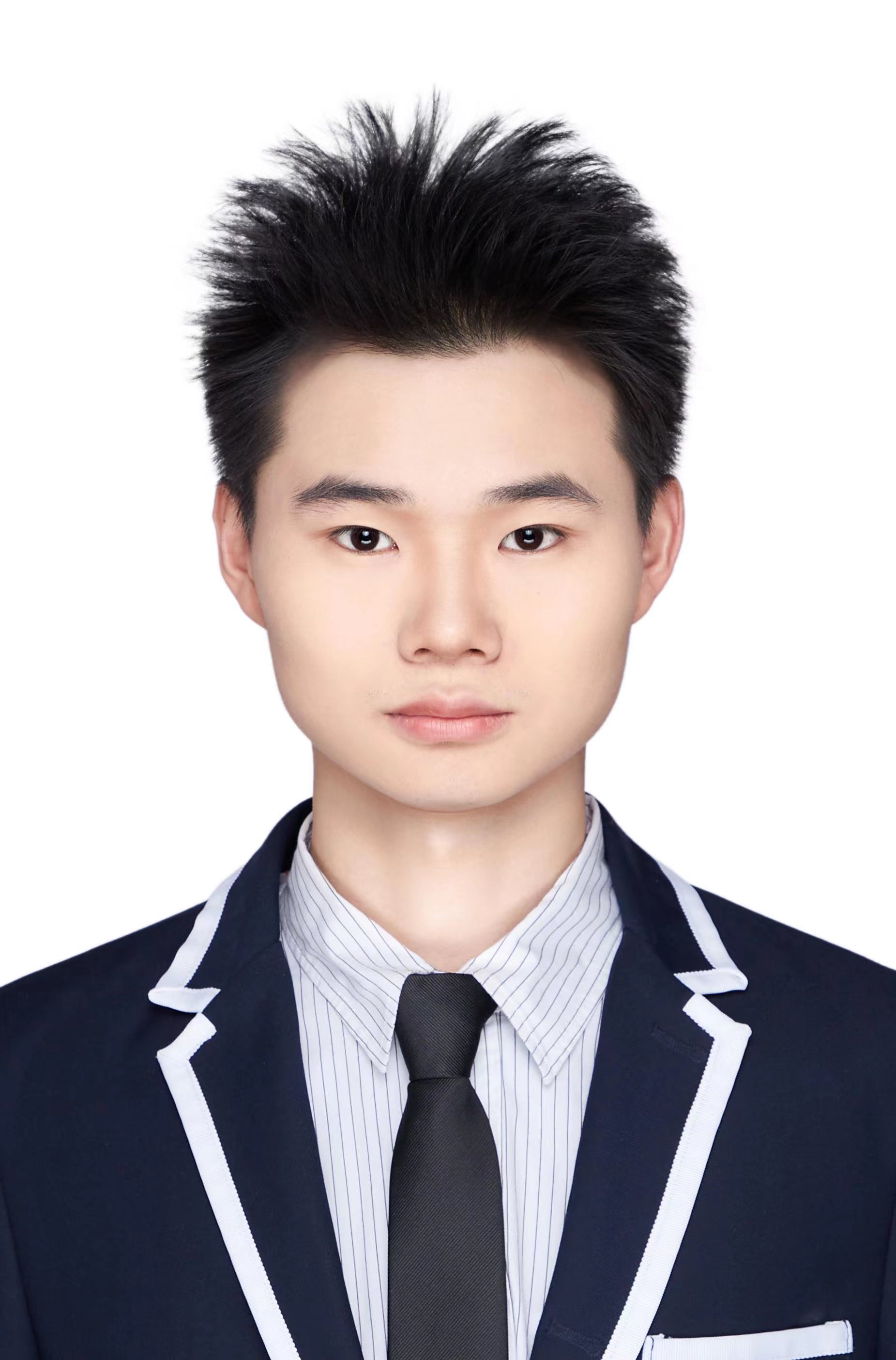}}]{Fang Jiahao} received the B.S. degree in Microelectronics Science and Engineering from Tsinghua University, Beijing, China, in 2021.
    
Currently, he is pursuing the M.S. degree at the School of Integrated Circuits, Tsinghua University, with a research focus on AI large models and hardware architecture. 
\end{IEEEbiography}
\vspace{-20 pt}

\begin{IEEEbiography}[{\includegraphics[width=1in,height=1.25in,clip,keepaspectratio]{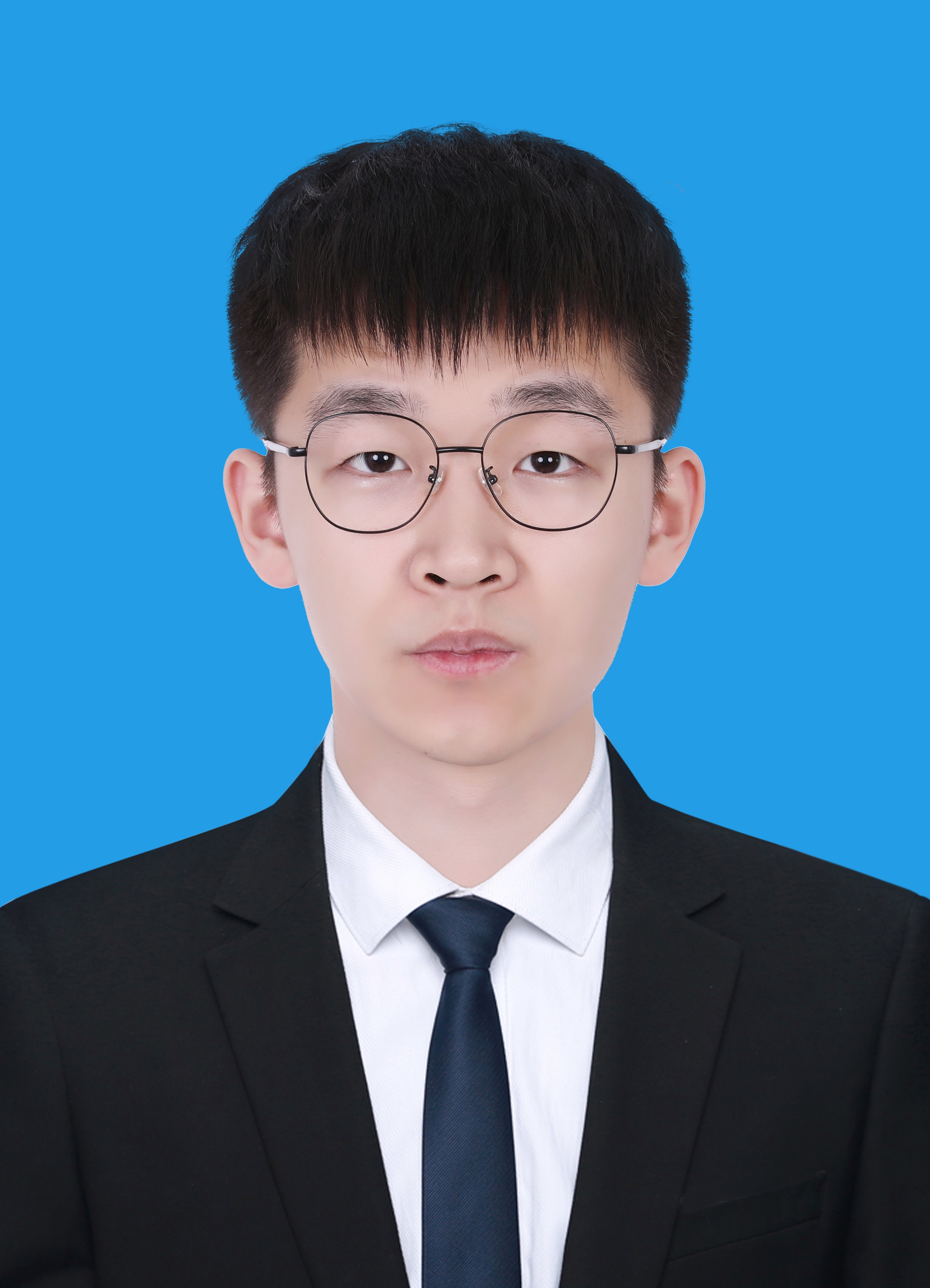}}]{Haoran Shang} received the B.S. degree in communication engineering from Beihang University, Beijing China, in 2023.
    
He is currently a graduate student with the School of Integrated Circuits, Tsinghua University. His research interests include network in chip, on-package interconnect.
\end{IEEEbiography}
\vspace{-20 pt}

\begin{IEEEbiography}[{\includegraphics[width=1in,height=1.25in,clip,keepaspectratio]{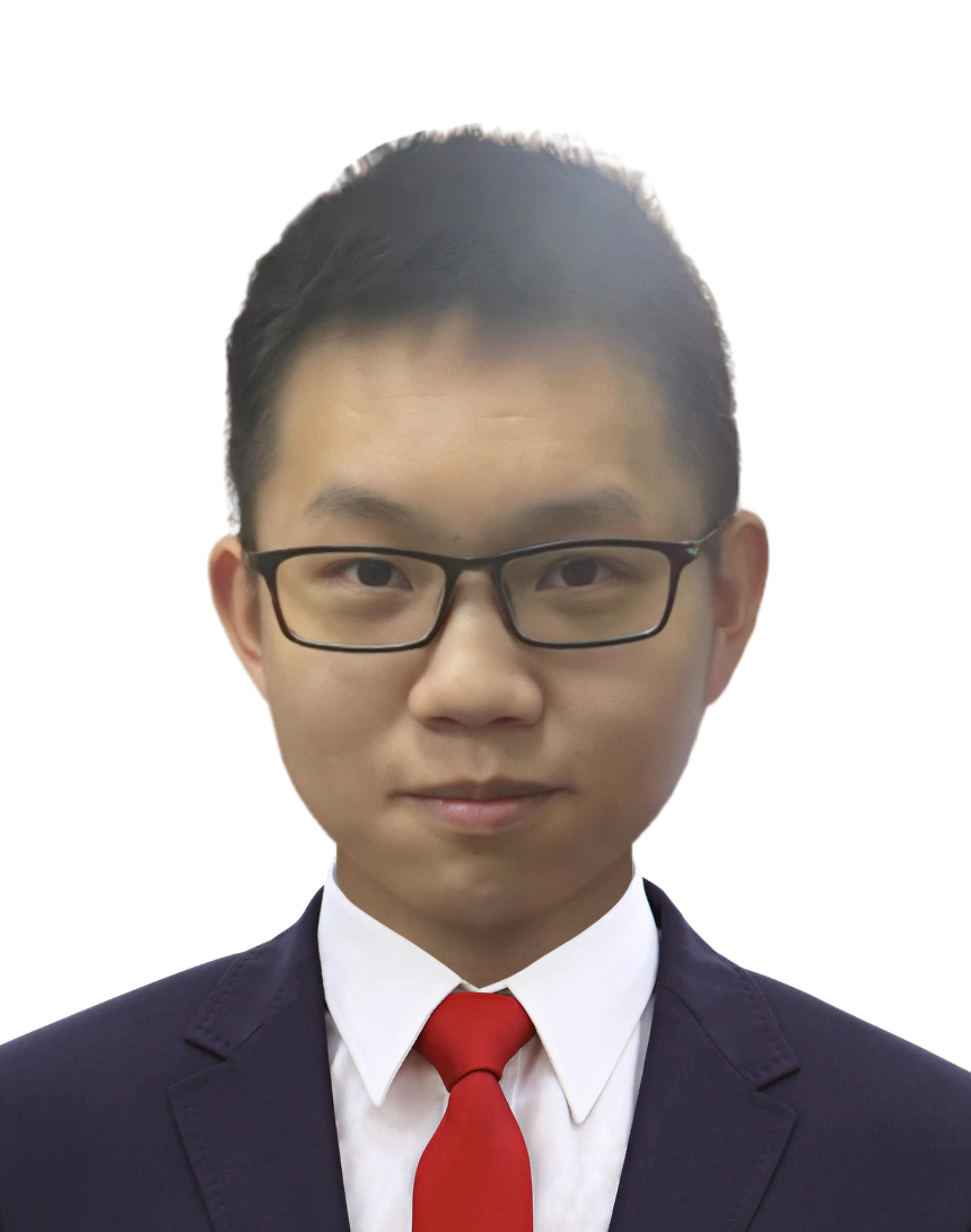}}]{Xinru Tang} received the B.S. degree in Integrated Chip Design and Integrated System from Huazhong University of Science and Technology, Hubei, China, in 2023. He is currently working toward the Ph.D. degree in School of Integrated Circuits, Tsinghua University. His current research interests include microarchitecture of spatial accelerator, AI processor and large-scaling chip design.
\end{IEEEbiography}
\vspace{-20 pt}

\begin{IEEEbiography}[{\includegraphics[width=1in,height=1.25in,clip,keepaspectratio]{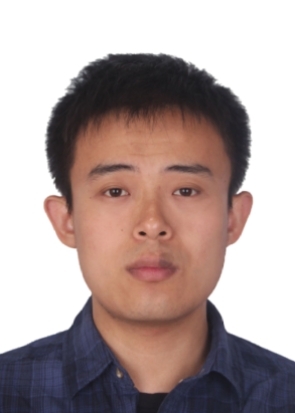}}]{Xu Dai} received the B.S. in microelectronics from Xidian University, Xi'an, China, in 2013, and the M.S. degree in electronic science and technology from Tsinghua University, Beijing, China, in 2016.

He is currently working with the Shanghai Artificial Intelligence Laboratory, Shanghai, China. His research interests include AI compiler, AI system and deep learning.
\end{IEEEbiography}
\vspace{-20 pt}

\begin{IEEEbiography}[{\includegraphics[width=1in,height=1.25in,clip,keepaspectratio]{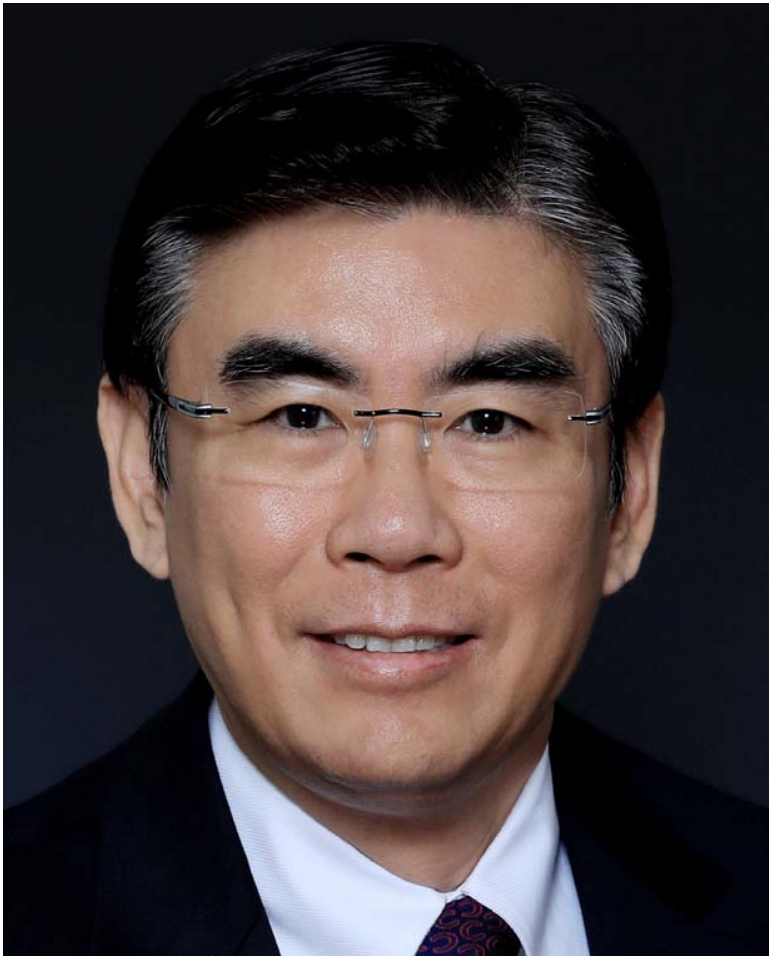}}]{Shaojun Wei} (Fellow, IEEE) was born in Beijing, China, in 1958. He received the Ph.D. degree from the Faculte Polytechnique de Mons, Mons, Belgium, in 1991.

He became a Professor at the Institute of Microelectronics, Tsinghua University, Beijing, China, in 1995. His main research interests include VLSI SoC design, electronic design automation (EDA) methodology, and communication application-specific integrated circuit (ASIC) design.
    
Dr. Wei is a Senior Member of the Chinese Institute of Electronics (CIE).
\end{IEEEbiography}
\vspace{-20 pt}

\begin{IEEEbiography}[{\includegraphics[width=1in,height=1.25in,clip,keepaspectratio]{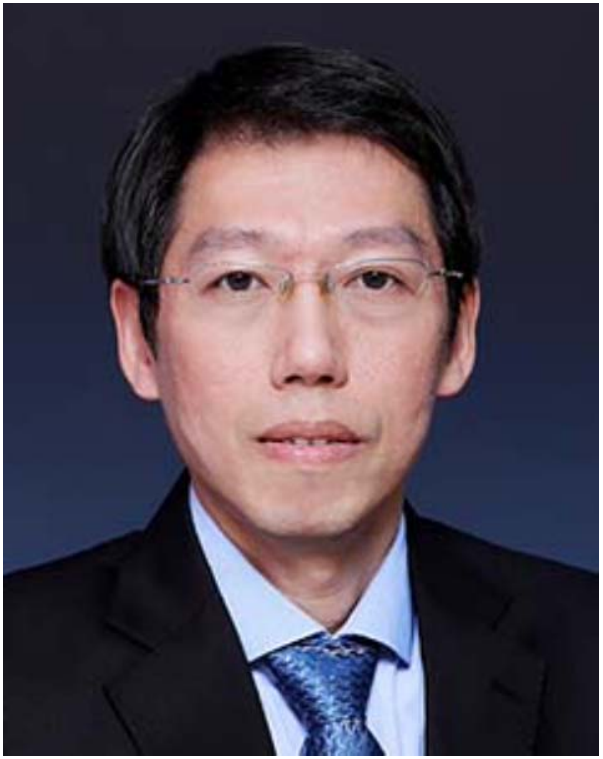}}]{Shouyi Yin} (Member, IEEE) received the B.S., M.S., and Ph.D. degrees in electronic engineering from Tsinghua University, Beijing, China, in 2000, 2002, and 2005, respectively.

He was a Research Associate with Imperial College London, London, U.K. He is currently a Full Professor and the Vice-Director of the School of Integrated Circuits, Tsinghua University. He has published more than 100 journal articles and more than 50 conference papers. His research interests include reconfigurable computing, AI processors, and high-level synthesis.
   
Dr. Yin has served as a Technical Program Committee Member of the top very-large-scale integration (VLSI) and electronic design automation (EDA) conferences, such as the Asian Solid-State Circuits Conference (A-SSCC), the IEEE/ACM International Symposium on Microarchitecture (MICRO), the Design Automation Conference (DAC), the International Conference on Computer-Aided Design (ICCAD), and the Asia and South Pacific Design Automation Conference (ASP-DAC). He is also an Associate Editor of IEEE Transactions on Circuits and Systems—I: Regular Papers, ACM Transactions on Reconfigurable Technology and Systems (TRETS), and Integration, the VLSI Journal.
\end{IEEEbiography}
\vspace{-20 pt}


\vfill

\end{document}